\documentclass[aps,pra, twocolumn,superscriptaddress,longbibliography]{revtex4-2}

\usepackage{graphicx}
\usepackage{float}

\usepackage[dvipsnames]{xcolor}
\usepackage{derivative}
\usepackage{xcolor}
\usepackage[margin=0.68in]{geometry}
\usepackage{siunitx}
\usepackage{soul}
\usepackage{minitoc}
\usepackage{color}
\usepackage{amssymb}
\usepackage{amsmath, scalerel, breqn}
\usepackage{colortbl}
\usepackage{bbm}
\usepackage{physics}
\usepackage{bbold}
\usepackage{booktabs, multirow}
\usepackage{changepage,threeparttable}
\usepackage{makecell}

\definecolor{dark-red}{rgb}{0.84,0.15,0.15}
\definecolor{dark-blue}{rgb}{0.15,0.15,0.4}
\definecolor{medium-blue}{rgb}{0,0,0.5}
\usepackage{hyperref}
\hypersetup{
    colorlinks, linkcolor={dark-red},
    citecolor={dark-blue}, urlcolor={medium-blue}
}

\makeatletter
\def\@fnsymbol#1{\ensuremath{\ifcase#1\or *\or \dagger\or \dagger\or
   \mathsection\or \mathparagraph\or \|\or **\or \dagger\dagger
   \or \ddagger\ddagger \else\@ctrerr\fi}}
\makeatother

\renewcommand{\thetable}{\arabic{table}}

\begin{document}
\doparttoc
\faketableofcontents 
\part{}


\title{Dynamics of magnetization at infinite temperature in a Heisenberg spin chain}
\author{Google Quantum AI and Collaborators}

\begin{abstract} 
Understanding universal aspects of quantum dynamics is an unresolved problem in statistical mechanics. In particular, the spin dynamics of the 1D Heisenberg model were conjectured to belong to the Kardar-Parisi-Zhang (KPZ) universality class based on the scaling of the infinite-temperature spin-spin correlation function. In a chain of 46 superconducting qubits, we study the probability distribution, $P(\mathcal{M})$, of the magnetization transferred across the chain's center. The first two moments of $P(\mathcal{M})$ show superdiffusive behavior, a hallmark of KPZ universality. However, the third and fourth moments rule out the KPZ conjecture and allow for evaluating other theories. Our results highlight the importance of studying higher moments in determining dynamic universality classes and provide insights into universal behavior in quantum systems.
\end{abstract}

\maketitle 

In statistical physics, the notion of universality is a powerful assertion; it implies that systems with entirely different microscopic interactions can share the same emergent macroscopic description because they have certain basic physical properties in common. It is a triumph of universality that, for instance, the Ising model is crucial to our understanding of the zero-temperature phase transitions in a wide class of systems\,\cite{KADANOFF1990,SachdevBook}. The basic ingredients commonly affecting universality classes are the collective behavior of constituent elements, symmetries, conservation laws, and dimensionality, as described by the renormalization group\,(RG) theory\,\cite{Geza}. In contrast to rather well-understood low-temperature universality classes, which are determined by ground-state physics, we have limited knowledge of the universality classification of dynamical phases of matter at finite temperatures, where contributions from the entire spectrum must be considered\,\cite{PolkovnikovRMP2011,gogolin2016equilibration}.

\vspace{0.5mm}
It has been observed in several dynamical systems that the long-time behavior permits a few-parameter hydrodynamical description, suggesting the existence of universality\,\cite{Zwierlein2011,erne2018universal,joshi2022observing, zu2021emergent, wei2022quantum,fontaine2021observation}. The emergence of a hydrodynamical description relies on reaching local, and subsequently, global equilibrium\,\cite{MitraPRA2014,Vir_PRB2018}. This fate is less certain in systems with an extensive set of conserved quantities, i.e., integrable systems, which are known to evade thermalization, and their universal behaviors are discussed in the framework of generalized hydrodynamics\,\cite{OlallaPRX2016,Bertini_PRL_2016,DoyonSciPostPhys,BulchandaniPRL2017,de2018hydrodynamic}. 

Distinct microscopic models or dynamics  belong to the same universality class if they share a single scale-invariant limit under a RG flow\,\cite{KADANOFF1990,Geza}. A universality class is commonly characterized by scaling exponents and scaling functions, and it is rather implausible to extract them all experimentally. Therefore, experiments, e.g., on quantum processors, cannot prove that a set of observed dynamics belongs to a given class, but they can falsify a universality conjecture\,\cite{Popper} by examining its predictions. They can also probe numerically and theoretically challenging regions of the parameter space, which has proven advantageous for studying universal behaviors\,\cite{Zwierlein2011,erne2018universal,joshi2022observing, zu2021emergent, wei2022quantum,fontaine2021observation}. 

Superconducting quantum processors offer high wavefunction sampling rates, which enabled them to show quantum advantage over classical computers in sampling tasks\,\cite{supremacy2019,Pan_PRL_2021}. On these processors one can go beyond mean expectation values and provide ``snapshots'' of an observable, which allows for measuring quantum fluctuations and the probability distribution of the observable. The capability of collecting full counting statistics could have fundamental consequences for our understanding of dynamical universalities. In particular, it is commonly assumed that the scaling functions and exponents of the first few moments uniquely determine a universality class, and there have not yet been any instances where the higher moments of an observable have led to a different classification.

\begin{figure*}[t!]
    \centering
    \includegraphics[width=0.98\textwidth]{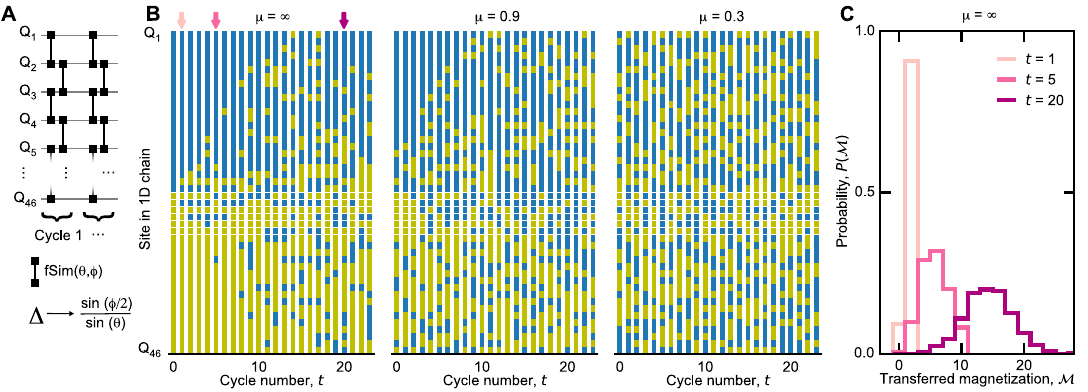}
    \caption{\small{\textbf{Domain wall relaxation in the Heisenberg XXZ spin chain}. (\textbf{A}) Schematic of the unitary gate sequence used in this work, where fSim gates are applied in a Floquet scheme on a 1D chain of $N_Q=46$ qubits. (\textbf{B}) Relaxation dynamics as a function of site and cycle number for $\mu = \infty , 0.9$, and $0.3$ for initially prepared domain-wall states with $2\langle S^z\rangle=\pm \tanh \mu$. Blue and yellow squares correspond to occupied and unoccupied sites, respectively, in random instances of the experiment. The fSim angles are chosen to be $(\theta, \phi) = (0.4\pi, 0.8\pi)$, corresponding to $\Delta = 1$. (\textbf{C}) Histogram showing the probability distribution of transferred magnetization after $t=1$, $5$ and $20$ cycles (arrows in \textbf{B}) for $\mu =\infty$.}}
    \label{fig:Fig1}
\end{figure*}

\section{Spin dynamics in Heisenberg XXZ spin chains}

Spin dynamics in the one-dimensional (1D) XXZ model have been the subject of numerous recent studies\,\cite{Znidaric2011,Bertini_PRL_2016,ljubotina2017spin,stephan2017return,sanchez2018anomalous,Ilievski_PRL_2018,Sarang_PRL_2019,DeNardisPRL2019,gopalakrishnan2019anomalous,de2020superdiffusion,MoorePRB2020,Dupont_PRL_2021,IlievskiPhysRevX2021, scheie2021detection,jepsen2020spin,wei2022quantum}. This integrable model describes nearest-neighbor exchange interactions between spin-$1/2$ particles, with the Hamiltonian\,\cite{Heisenberg}  
\begin{equation}
\label{eq:HXXZ}
 H=\sum_i \left(  S^x_i  S^x_{i+1}+  S^y_i  S^y_{i+1} + \Delta  S^z_i  S^z_{i+1} \right),
\vspace{-2.2mm}
\end{equation}
where $S^x$, $S^y$, and $S^z$ are spin-1/2 operators, and $\Delta$ is the anisotropy parameter. When $\Delta=1$, this system is the Heisenberg model, a paradigmatic model of quantum magnetism that possesses a global SU$(2)$ rotational symmetry. The spin dynamics in the Heisenberg model exhibit characteristics consistent with the Kardar-Parisi-Zhang (KPZ) universality class, which was originally introduced to describe the stochastic, nonlinear dynamics of driven interfaces and has proven to apply to a wide range of classical systems \,\cite{ljubotina2017spin,stephan2017return,sanchez2018anomalous,Ilievski_PRL_2018,Sarang_PRL_2019,DeNardisPRL2019,gopalakrishnan2019anomalous,de2020superdiffusion,MoorePRB2020,Dupont_PRL_2021,IlievskiPhysRevX2021, scheie2021detection,wei2022quantum}. The KPZ-like behavior of the spin dynamics is surprising because of the absence of stochasticity and nonlinearity in the Heisenberg model.

\vspace{0.6mm}
In a 1D chain of $N_Q=46$ superconducting qubits, we simulate this spin model by periodic (Floquet) application of high-fidelity 2-qubit unitary fSim$(\theta, \phi)$ gates \,(Fig.\,1A, Refs.~\cite{Neill2021accurately, supplement}, Fig.~S7). Here, $\theta$ sets the amplitude of hopping between adjacent qubit lattice sites, and $\phi$ is the conditional phase angle imparted when two spin excitations are adjacent to each other. Within each cycle, two-qubit fSim$(\theta, \phi)$ gates are applied between all neighboring pairs in the chain, resulting in the cycle unitary: 
\begin{equation}
\label{eq:HfSim}
U_F= \prod_{\text{even bonds}} \text{fSim}(\theta, \phi) \prod_{\text{odd bonds}} \text{fSim}(\theta, \phi). 
\end{equation}
\noindent In the limit $\theta,\,\phi\,\rightarrow\,0$, $U_F$ is the Trotter–Suzuki expansion of the XXZ Hamiltonian, Eq.~\ref{eq:HXXZ}, with $\Delta = \sin(\phi/2)/\sin(\theta)$. Away from this limit, there is no unique Hamiltonian associated with $U_F$, but Eqs.~\ref{eq:HXXZ} and \ref{eq:HfSim} still share symmetries and are both integrable by the Bethe ansatz\,\cite{Bethe1931,Essler_PRL_2013,ProsenTrotter, Integrability2, Ljubotina_Ballistic_2019}. The conjecture that the late-time dynamics are described by KPZ applies equally to the Floquet system\cite{Ljubotina_PRL_2019}, so we make no attempt to be in the small-angle limit, instead favoring large angles for faster dynamics.

\begin{figure*}[t!]
    \centering
    \includegraphics{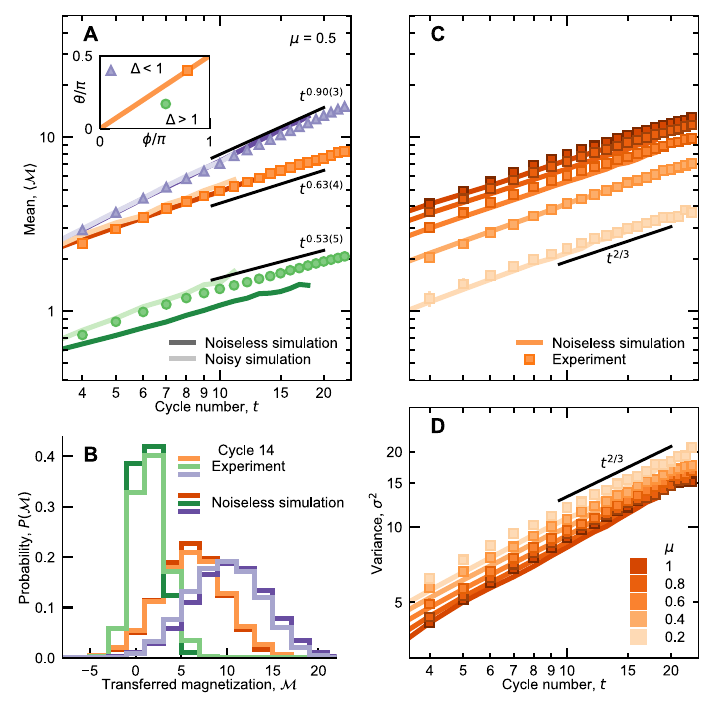}
    \caption{\small{\textbf{Mean and variance in various transport regimes}. (\textbf{A}) Mean of transferred magnetization $\langle\mathcal{M}(t)\rangle$ as a function of cycle number for initial states with $\mu =0.5$ and for $\Delta=0.16$ (purple triangles), 1 (orange squares), 1.6 (green circles). Light and dark curves show simulations with and without noise, respectively. The $\langle\mathcal{M}(t)\rangle$ can be fit to $t^{1/z}$ and gives $z=1.12 \pm 0.04 $ in the ballistic, $z=1.6 \pm 0.1 $ in the superdiffusive, and $z=1.9 \pm 0.2 $ in the diffusive regime. The inset illustrates three different regimes characterized by $\Delta=\sin(\phi/2)/\sin(\theta)$ with the orange line being the isotropic Heisenberg limit. (\textbf{B}) Histogram showing the probability distribution of measured $\mathcal{M}$ for values of $\Delta$ studied in \textbf{A} at cycle 14. Light and dark lines show experimental data and noiseless simulation results, respectively. (\textbf{C},\textbf{D}) Mean and variance of $\mathcal{M}$ for $\Delta=1$ and $0.2 \leq \mu\leq1$ (brighter to darker squares). With increasing $\mu$, the mean increases, whereas the variance decreases.}}
    \label{fig:Fig2}
\end{figure*}

To study dynamics under the unitary evolution (Eq.~\ref{eq:HfSim}), we generate domain-wall initial states with an adjustable contrast parameter $\mu$ (Fig.\,1B). Specifically, we initialize the chain in a set of product states such that the left and right halves have average magnetization $\pm \tanh(\mu)$, respectively:

\begin{equation}
\label{eq:mu}
\rho(t=0) \propto (e^{2\mu S^z})^{\otimes N_Q/2}\,  \otimes \,( e ^{-2\mu S^z}) ^{\otimes N_Q/2}. 
\vspace{1mm}
\end{equation}
\noindent
When $\mu \rightarrow \infty$, the system approaches a pure domain-wall state with the two sides fully magnetized in opposite directions. Only when $\mu=0$, the initial state is an infinite-temperature thermal state that preserves SU(2) symmetry. When $ \mu \ne 0$, the magnetization is preferentially along the $z$-axis, breaking the SU$(2)$ rotational symmetry of the Heisenberg model.

\vspace{0.6mm}  
A natural measure of spin transport is the total transferred magnetization, ${\mathcal{M}}(t)$, defined as twice the net number of excitations that have crossed the middle of the chain after $t$ cycles. In our experiment, we sample over initial bitstring states with probabilities given by Eq.~\ref{eq:mu}. For each initial state, we prepare the qubits in that state and then apply $t$ cycles of fSim gates. Let $N_{R,1}(b)$ be the number of excitations (``1"s) in the right half of bitstring $b$. The transferred magnetization $\mathcal{M}$ is the stochastic variable defined by
\begin{equation}
    \mathcal{M}(t)/2 = N_{R,1}(b_t) - N_{R,1}(b_i),
\end{equation}
where $b_i$ is the initial bitstring, sampled from Eq.~\ref{eq:mu}, and $b_t$ is the associated final bitstring sampled at $t$. For example, if the initial bitstring is 111010 and the final bitstring is 110110, then the transferred magnetization is 2. Because the dynamics are number-conserving, the transferred magnetization is also the net number of zeros that have crossed from the right to the left. Repeating the experiment many times, we construct the probability distribution of $\mathcal M$, $P(\mathcal{M})$. In the case $\mu=0$, the initial state and the dynamics both have a mirror symmetry, so for each initial bitstring that is studied experimenally, we also include the reflection of that bitstring in our analysis, using the same experimental data, which effectively symmetrizes $P(\mathcal{M})$.

Figure 1B shows measurement instances for three values of $\mu$. The left column in each panel shows an instance of the initial state for the given $\mu$, and the subsequent columns show typical bitstrings evolved from that state. As excitations (spin flips) propagate through the chain, smaller domains become more probable. 
In Fig.\,1C, we show histograms of $\mathcal{M}$ at different times, starting in a pure ($\mu=\infty$) domain wall. Owing to locality of the circuit, $|\mathcal M(t)|$ is upper-bounded by $2\,t$. Consequently, the distribution is narrow and centered around a small value at $t=1$, because only a few excitations have crossed the middle of the chain, and becomes wider at later times. 

\section{Mean and variance of transferred magnetization}
 
In the context of spin transport, the first and second (variance) moments of ${\mathcal{M}}$ have been extensively studied both theoretically and experimentally\,\cite{Bertini_PRL_2016,ljubotina2017spin,stephan2017return,sanchez2018anomalous,Ilievski_PRL_2018,Ljubotina_Ballistic_2019,Sarang_PRL_2019,DeNardisPRL2019,gopalakrishnan2019anomalous,de2020superdiffusion,MoorePRB2020,Dupont_PRL_2021,IlievskiPhysRevX2021, scheie2021detection,wei2022quantum,keenan2022evidence}. Taking advantage of our tunable fSim gates, we explore how these two moments depend on the anisotropy parameter, $\Delta$. Figure 2A shows the mean of ${\mathcal{M}}$ over time for values of $\Delta$ equal to 0.16 (purple), 1 (orange) and 1.6 (green), and an initial domain wall height of $\mu=0.5$. We observe markedly different scaling behaviors in the three regimes. Eliminating the initial transient cycles, we fit a power law, $\langle \mathcal{M} \rangle \sim t^{1/z}$, to the data over cycles 10--23 and extract scaling exponents of $z =1.12 \pm 0.04 $, $z = 1.6\pm 0.1$ and $z = 1.9\pm 0.2 $, respectively. These are in close agreement with theoretical predictions for the ballistic ($z=1$)\cite{Zotos99}, superdiffusive ($z=3/2$)\cite{ljubotina2017spin}, and diffusive ($z=2$)\cite{SachdevDamle97,DamleSachdev98} behaviors, respectively (although $z$ is predicted and observed to depend on $\mu$; see Fig.~S12 \cite{supplement}). Observation of superdiffusive propagation for isotropic interactions ($\Delta=1$), measured here and also in other works\,\cite{Znidaric2011,ljubotina2017spin,stephan2017return,sanchez2018anomalous,Ilievski_PRL_2018,DeNardisPRL2019,Sarang_PRL_2019,gopalakrishnan2019anomalous,MoorePRB2020,de2020superdiffusion,Dupont_PRL_2021,IlievskiPhysRevX2021, scheie2021detection,wei2022quantum,keenan2022evidence}, has been interpreted as a signature of the KPZ universality class.

Numerical simulations of these domain-wall dynamics are shown with solid dark lines in Fig.\,2A. A variety of numerical simulations often rely on approximation schemes, which could lead to inaccurate results. In contrast, here we perform exact statevector sampling up to cycle 18 without any approximations. This is achieved by taking advantage of the fact that $\langle\,\mathcal{M}(t)\,\rangle$ depends only on the spins within the light cone of width $2t$, and can thus be determined exactly by simulating shorter chains. This simplification also allows for arriving at analytical results for all moments of $\mathcal{M}$\, at early cycles. Nevertheless, the computational cost grows exponentially, and with the resources used here, the simulations at cycles 14, 16, and 18 take about 1, 2, and 14 hours, respectively\,(see Fig.~S9 \cite{supplement}). In contrast, the quantum simulator allows us to reach 23 cycles, primarily limited by the relaxation of the qubits to the $|0\rangle$ state, which necessitates an exponentially large sampling overhead in the absence of error correction.

Importantly, the slight discrepancies between the observed and predicted exponents are also seen in the exact statevector simulations up to $t=18$ cycles (colored lines), which agree almost perfectly with the experimental data for $\Delta < 1$ and $\Delta = 1$ and indicate that the deviations of the exponents from 1 and $2/3$, respectively, primarily stem from finite-time and large-$\mu$ effects rather than from experimental imperfections. Indeed, the exact scaling exponents are only expected in the long-time limit and as $\mu\to0$. Furthermore, by simulating the effects of noise in our system\,(lighter lines in Fig.\,2A; referred to as ``noisy simulations" elsewhere), we see that our noise model explains why we find near-perfect agreement between noiseless simulation and experiment in the $\Delta<1$ and $\Delta=1$ cases and not in the $\Delta>1$ case, where $\mathcal{M}$ is smaller. This effect is also noticeable in the magnetization transfer distributions in Fig.\,2B. Because the distribution is narrower in the $\Delta > 1$ case, the noise has a larger effect on the shape of the distribution here than for the other two values of $\Delta$. The error in this case is found to be predominantly caused by combined occurrences of decay to the $|0\rangle$ state and $0\rightarrow 1$ readout errors, which are not eliminated by post-selection. By including this effect in the simulation, we find good agreement in all three regimes (Figs.~2A, S5, S15, S16) \cite{supplement}. 

\begin{figure*}[t!]
    \centering
    \includegraphics{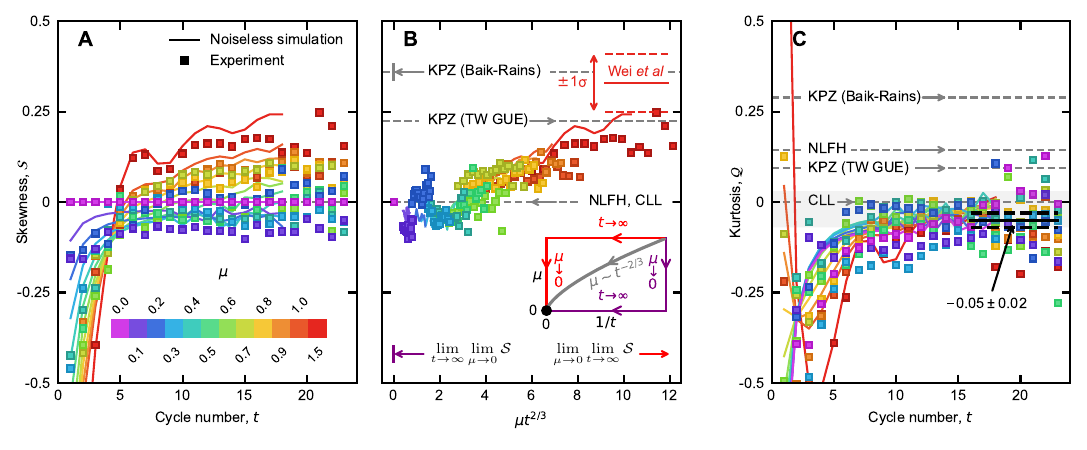}
    \caption{\small{\textbf{Skewness and excess kurtosis of transferred magnetization}. Experimental data and noiseless simulation results are shown with squares and lines, respectively. (\textbf{A}) Skewness of transferred magnetization distribution $\mathcal{S}$ as a function of $t$, for $\Delta=1$ and various $ \mu \in [0,1.5]$.  We symmetrize the $\mu=0$ probability distribution, after which the skewness is exactly 0. (\textbf{B}) The same as A, but with the $x$-axis re-scaled as $\mu\,t^{2/3}$ and excluding data points for which $t<8$. The collapse of the skewness under rescaling is explored using the noiseless simulation data in Fig.~S10 \cite{supplement}. Dashed horizontal gray lines indicate  predictions based on the KPZ universality class (TW GUE: Tracy-Widom Gaussian-unitary-ensemble), nonlinear fluctuating hydrodynamics (NLFH) model\,\cite{denardis2023nonlinear}, and a classical Landau-Lifshitz (CLL) model\,\cite{krajnik2023universal}. The red lines marked ``Wei \textit{et al}" show $\mathcal{S}$ measured in Ref.~\cite{wei2022quantum} (for $\mu=1.5$) and the $1\sigma$ confidence interval. The {\bf inset} to B illustrates different ways of taking the late-time and small-$\mu$ limits that are considered here. ({\bf C}) Kurtosis of transferred magnetization. The horizontal lines are the theoretical predictions from the same models shown in panel B. The experimental kurtosis, averaged over cycles 16--23, is $-0.05 \pm 0.02$. The kurtosis data do not exhibit a collapse, as expected; the kurtosis cannot be a function of $\mu t^\gamma$ for any $\gamma$ because, unlike the skewness, it has time dependence even when $\mu=0$.}}
    \label{fig:Fig3}
\end{figure*}

\vspace{0.6mm} 
Superdiffusive transport, $\langle\mathcal{M}\rangle$\,$\sim$\,$ t^{2/3}$, at $\Delta = 1$ is a characteristic of systems within the KPZ universality class. Moreover, numerical studies found that the spin-spin correlation function coincides with the KPZ scaling function\,\cite{Ljubotina_PRL_2019,Prahofer2004}, which has led to the conjecture that near-equilibrium spin transport in the Heisenberg model belongs to the KPZ universality class\,\cite{Ljubotina_PRL_2019,KIP2020,Norm_PRL_2022}. This universality class is associated with the classical nonlinear stochastic KPZ equation $\partial h/\partial t = \nu \nabla^2 h + \lambda (\nabla h)^2 + \eta(x,t)$, which was originally introduced\,\cite{KPZ} to describe the dynamics of driven interfaces as a height field $h(x,t)$, where $\nu$, $\lambda$, $\eta$ set the strength of the smoothening diffusion, roughening nonlinear growth, and stochasticity terms, respectively. The conjecture asserts that at late times the magnetization profile behaves similarly to $\partial h(x,t) / \partial x $. Consequently,
\begin{equation}
\label{eq:KPZ_conjecture}
\lim_{\mu \rightarrow 0} \mathcal{M} (t)  \,\, \longleftrightarrow  \,\,  2h(0,t)- h(-\infty,t)- h(\infty,t).
\end{equation}

To further examine the universality class of the Heisenberg spin dynamics, two aspects are of particular importance. First, because the universal behavior is expected to depend on whether the system is in equilibrium, it is essential to measure the dependence on $\mu$. Second, although the scaling exponent of the mean is consistent with the KPZ universality class, further insights can be gained by examining higher moments (the ``full counting statistics'') of $P(\mathcal{M})$. Owing to the reduced signal-to-noise ratio, measuring higher moments at small $\mu$ is experimentally challenging. We utilize our fast sampling capability to measure $P(\mathcal{M})$ as a function of $\mu$ and $t$\,(Refs.\,\cite{Krajnik_PRL_2022,gopalakrishnan2022theory}). Figure 2, panels C and D, shows the temporal evolution of the mean and variance $\mathcal{M}$ for various values of $\mu$ ranging from 0.2 to 1. We emphasize that the KPZ-conjecture is only expected to apply for small $\mu$, in which case we find that the dynamical exponents of both the mean and the variance are close to 3/2. For larger $\mu$, the dynamical exponent of the mean approaches 5/3, consistent with recent numerical results\,\cite{ljubotina2017spin, supplement} (Fig.~S12), and confirming that small $\mu$ is required for potentially recovering KPZ dynamics.

\section{Higher-order moments}

\vspace{0.6mm}
Next, we extract the skewness $\mathcal{S}$ and kurtosis $\mathcal{Q}$ of $P(\mathcal{M})$, 
\begin{equation}
\mathcal{S} = \frac{ \alpha_3 }{\alpha_2^{3/2}},\,\hspace{3mm} \mathcal{Q} = \frac{ \alpha_4 }{\alpha_2^{2}} - 3,  
\end{equation}
\noindent
where $\alpha_k=\left\langle (\mathcal{M}-\langle\mathcal{M}\rangle)^k\right\rangle$ is the $k^{\mathrm{th}}$ moment. In Fig.\,3A, we show the temporal dependence of $\mathcal{S}$ for $\mu$ ranging from 1.5 to 0.1. Consistent with Ref.~\onlinecite{wei2022quantum}, $\mathcal{S}$ goes up to about 0.25 for $\mu >1$. However, as $\mu$ is reduced towards the infinite-temperature equilibrium point, we observe that $\mathcal S$ goes to zero. Figure 3C shows that for later cycles, the initial strong time dependence of $\mathcal{Q}$ weakens. By averaging over cycles 16 to 23, we obtain a kurtosis of $-0.05\pm 0.02$. Statistical error bars on the individual data points are shown in Fig.\,S8 \cite{supplement}.

In order to test the KPZ universality conjecture, one needs to study the infinite-time ($t \rightarrow \infty$) and near-equilibrium ($ \mu \rightarrow 0$) limits. These limits are experimentally inaccessible. However, if there exists a function $f(\mu, t)$ such that the moments are functions of $f(\mu,t)$, then one may be able to extrapolate measured values at finite $\mu$ and $t$ to these unattainable limits. We empirically find that the zero crossing of $\mathcal{S}$ scales as $t_0\sim\mu ^{-1.49}$ (Fig.~S10) \cite{supplement}, suggesting that $\mathcal{S}$ may be a function of $\mu t^{2/3}$. Indeed, after excluding the initial transient behavior, $\mathcal{S}$ does appear to be a single-valued function of $\mu t^{2/3}$ (Figs.\,3B, S10) \cite{supplement}.

KPZ has been conjectured to apply to high-temperature thermal states at late times, corresponding to taking $\mu\to0$ first and then $t\to\infty$\,(Ref.\,\cite{Ljubotina_PRL_2019}). In this case, $P(\mathcal{M})$ should become the Baik-Rains distribution \cite{PrahoferSpohn2000}. However, this distribution is skewed (Table \ref{tab:Table1}), whereas symmetry dictates that $\mathcal{S} = 0$, consistent with the trend observed in the experimental data (Fig.~3, A and B).

One might also search for KPZ universality away from $\mu=0$, corresponding to a different order in taking these noncommuting limits. When taking $t\to\infty$ first, the appropriate probability distribution to compare $P(\mathcal{M})$ against is the Tracy Widom (TW) distribution\,\cite{PrahoferSpohn2000}, which has $\mathcal{S}$ of about 0.22. This order of limits corresponds to large $\mu t^{2/3}$ in Fig.\,3B, where we indeed find $\mathcal{S}$ consistent with this distribution; averaging over the four rightmost points in Figure 3B, we find $\mathcal{S} = 0.18 \pm 0.02$, consistent to two standard deviations with the TW GUE value, as well as with an earlier experiment, which found $\mathcal{S} = 0.33\pm 0.08$ \cite{wei2022quantum}. However, $\mathcal{Q}$ of the TW GUE (Gaussian unitary ensemble) distribution is 0.09, whereas we find $\mathcal{Q} = -0.05\pm 0.02$. The emergence of KPZ dynamics in this order of limits is further ruled out by numerical and theoretical predictions that the dynamics become diffusive ($z=2$) at a late time that grows as $t\sim 1/\mu^3$ as $\mu\to0$ \,(Refs.\,\cite{Krajnik_PRL_2022, Gopalakrishnan_2023}).

One could consider taking the two limits simultaneously in a way that the dynamics do not become diffusive, e.g., by holding $\mu t^{2/3}$ constant (see Fig.~S18) \cite{supplement}. The correct distribution to compare against is TW GUE in this case as well. If we take the limit with $\mu t^{2/3}$ fixed at a large value, we find $\mathcal{S}$ consistent with TW GUE, but the measured  $\mathcal{Q}$ is still inconsistent with the TW GUE prediction of 0.09 (Fig.\,3C), ruling out KPZ dynamics on the timescales accessible in the experiment. Although it remains possible that KPZ dynamics will emerge at much later times (i.e., $\mathcal{Q}$ will increase to 0.09), we see neither evidence nor rationale for this.

An outstanding question is why only lower-point observables, such as the mean and variance of the transferred magnetization and the correlator studied by \cite{Ljubotina_PRL_2019}, seem to behave consistently with KPZ universality. Intriguingly, other systems have been identified that exhibit similar behavior. One such system is a nonlinear fluctuating hydrodynamic (NLFH) model with two coupled stochastic modes\,\cite{Vir2020_gauge, Sarang_PRB_2022,denardis2023nonlinear}, which predicts $\mathcal{S}=0$, consistent with the Heisenberg spin chain. However, it suggests $\mathcal{Q}=0.14$, differing from what we observe, perhaps because not all aspects of the model are universal. Another such system is the classical Landau-Lifshitz\,(CLL) magnet\,\cite{Krajnik_PRL_2022,krajnik2023universal, Manas1, Manas2}, which predicts $\mathcal{S}=0$ and a $\mathcal{Q}$ that is negative and close to zero at these time scales\,\cite{krajnik2023universal}. These are consistent with our experimental results. It is rather interesting that this classical system additionally exhibits dynamical behavior similar to the quantum spin chain with enhanced quantum fluctuations due to confinement\,\cite{Giamarchi}.

\definecolor{bubblegum}{rgb}{1.19, 0.83, 0.9}
\definecolor{teagreen}{rgb}{0.82, 0.97, 0.8}
\definecolor{Gray}{gray}{0.95}

\begin{table}
{\renewcommand{\arraystretch}{1.3}
{\setlength{\tabcolsep}{0.2mm}
\begin{tabular}
{|!{\hspace{1mm}}c<{\hspace{2mm}}|c<{\hspace{2mm}}c<{\hspace{2mm}}c<{\hspace{2mm}}c<{\hspace{1mm}}|} 
    \hline
    & $\langle\mathcal{M}\rangle$& $\sigma^2$&$\mathcal{S}$&$\mathcal{Q}$\\
    \hline
    
    Experiment&\cellcolor{Gray}$t^{2/3}$&\cellcolor{Gray}$\,\,t^{2/3}$&\cellcolor{Gray}$0$\,*&\cellcolor{Gray}$-0.05\pm 0.02$\\
    KPZ (Baik-Rains) \cite{PrahoferSpohn2000} & \cellcolor{teagreen}$t^{2/3}$& \cellcolor{teagreen}$\,\,t^{2/3}$&\cellcolor{bubblegum}\,0.36
    &\cellcolor{bubblegum}0.29\\
    
    NLFH \cite{denardis2023nonlinear} & \cellcolor{teagreen}$t^{2/3}$& \cellcolor{teagreen}\,\,$t^{2/3}$&\cellcolor{teagreen}0&\cellcolor{bubblegum}0.14\\
    CLL \cite{krajnik2023universal} & \cellcolor{teagreen}$t^{2/3}$& \cellcolor{teagreen}$\,\,t^{2/3}$&\cellcolor{teagreen}0&\cellcolor{teagreen} $\in [-0.07,0.03]$\\
    
    \hline
\end{tabular}}}
\caption{Comparison of the experimentally observed moments with the theoretically predicted values from various models. Here: $\langle\mathcal{M}\rangle$: mean, $\sigma^2$: variance, $\mathcal{S}$: skewness, $\mathcal{Q}$: kurtosis. The experimental kurtosis value is averaged over cycles $16-23$ and $\mu=0-0.4$, and the errors are computed using the jackknife method. See Table S2 for details regarding KPZ predictions \cite{supplement}. Green (red) entries indicate agreement (disagreement) between the data and theory. *The skewness is 0 due to symmetrization of our data.
}
\label{tab:Table1}
\end{table}

\section{Discussion and Outlook}

Studies of the universal aspects of quantum dynamics have attracted notable interest recently; accordingly, a complete classification of their universal properties is lacking. Our findings suggest these classifications could involve unanticipated subtleties. Our first result, also observed by others, is the superdiffusive transport characterized by $ \langle \mathcal{M} \rangle \sim t^{2/3} $, shown in Fig.\,2A. Although this anomalous diffusion is suggestive of the known KPZ universality classes, this classification is not compatible with our second finding---the vanishing of $\mathcal{S}$ and $\mathcal{Q}$ near equilibrium\,(Fig.\,3C). Despite the apparent consistency with the CLL model, a full understanding requires the development of a systematic spacetime RG framework that could establish the origin of the KPZ-like behavior starting from the microscopic dynamics of the Heisenberg model. Quantum processors have the potential to help with such RG studies (e.g. building on \cite{Vidal2008,evenbly2009algorithms})
For example, our results suggest that quantum entanglement (but not integrability) is irrelevant in the space-time RG sense.

Our observations are rooted in the interplay of integrability, quantum fluctuations, and symmetry and have proved to be challenging to describe using an effective quantum field theory. The observed discrepancies with KPZ predictions suggest that the infinite-temperature dynamics in the Heisenberg chain---if universal---belong to a yet-to-be-discovered dynamical universality class. 

\paragraph*{Acknowledgments:}
We acknowledge discussions with I.~Bloch, V.B.~Bulchandani, A.~Morningstar, and R.~Vasseur.
\paragraph*{Funding:}
V.K.~acknowledges support from the US Department of Energy, Office of Science, Basic Energy Sciences, under Early Career Award No.~DE-SC0021111, the Alfred P.~Sloan Foundation through a Sloan Research Fellowship, and the Packard Foundation through a Packard Fellowship in Science and Engineering. S.G., V.K., and T.P.~acknowledge the hospitality of the Kavli Institute for Theoretical Physics at the University of California, Santa Barbara (supported by NSF Grant PHY-1748958). R.S.~is supported by the Princeton Quantum Initiative Fellowship. T.P.~is supported by Program P1-0402 of Slovenian Research Agency (ARRS).


\paragraph*{Data and materials availability}
The data that support the findings in this study are available on Zenodo \cite{rosenberg_2024_10045874}. Code is available on ReCirq \cite{recirq}.


\onecolumngrid
\vspace{1em}

\begin{flushleft}
    {\hypertarget{authorlist}{${}^\dagger$}  \small Google Quantum AI and Collaborators}

    \bigskip

    \renewcommand{\author}[2]{#1\textsuperscript{\textrm{\scriptsize #2}}}
    \renewcommand{\affiliation}[2]{\textsuperscript{\textrm{\scriptsize #1} #2} \\}
    \newcommand{\corrauthora}[2]{#1$^{\textrm{\scriptsize #2}, \hyperlink{corra}{\ddagger}}$}
    \newcommand{\corrauthorb}[2]{#1$^{\textrm{\scriptsize #2}, \hyperlink{corrb}{\mathsection}}$}
    \newcommand{\xGoogle}{\affiliation{1}{Google Research, Mountain View, CA, USA}}

\newcommand{\xCornell}{\affiliation{2}{Department of Physics, Cornell University, Ithaca, NY, USA}}

\newcommand{\xPrinceton}{\affiliation{3}{Department of Physics, Princeton University, Princeton, NJ, USA}}

\newcommand{\xPCTS}{\affiliation{4}{Princeton Center for Theoretical Science, Princeton University, Princeton, NJ, USA}}

\newcommand{\xStanford}{\affiliation{5}{Department of Physics, Stanford University, Stanford, CA, USA}}

\newcommand{\xUCONN}{\affiliation{6}{Department of Physics, University of Connecticut, Storrs, CT}}

\newcommand{\xUMass}{\affiliation{7}{Department of Electrical and Computer Engineering, University of Massachusetts, Amherst, MA, USA}}

\newcommand{\xAU}{\affiliation{8}{Department of Electrical and Computer Engineering, Auburn University, Auburn, AL, USA}}

\newcommand{\xCQC}{\affiliation{9}{QSI, Faculty of Engineering \& Information Technology, University of Technology Sydney, NSW, Australia}}

\newcommand{\xUCR}{\affiliation{10}{Department of Electrical and Computer Engineering, University of California, Riverside, CA, USA}}

\newcommand{\xCU}{\affiliation{11}{Department of Chemistry, Columbia University, New York, NY, USA}}

\newcommand{\xUoCA}{\affiliation{12}{Department of Physics and Astronomy, University of California, Riverside, CA, USA}}

\newcommand{\xLjubljana}{\affiliation{13}{Faculty of Mathematics and Physics, University of Ljubljana, Ljubljana, Slovenia}}

\begin{footnotesize}

\newcommand{\Google}{1}
\newcommand{\Cornell}{2}
\newcommand{\Princeton}{3}
\newcommand{\PCTS}{4}
\newcommand{\Stanford}{5}
\newcommand{\UCONN}{6}
\newcommand{\UMass}{7}
\newcommand{\AU}{8}
\newcommand{\CQC}{9}
\newcommand{\UCR}{10}
\newcommand{\CU}{11}
\newcommand{\UoCA}{12}
\newcommand{\Ljubljana}{13}

\corrauthora{E. Rosenberg}{\Google,\! \Cornell},
\corrauthora{T. I. Andersen}{\Google},
\author{R. Samajdar}{\Princeton,\! \PCTS},
\author{A. Petukhov}{\Google},
\author{J. C. Hoke}{\Stanford},
\author{D. Abanin}{\Google},
\author{A. Bengtsson}{\Google},
\author{I.~K.~Drozdov}{\Google,\! \UCONN},
\author{C. Erickson}{\Google},
\author{P. V.~Klimov}{\Google},
\author{X. Mi}{\Google},
\author{A. Morvan}{\Google},
% ...
\corrauthorb{T. Prosen}{\Ljubljana},
\corrauthorb{P. Roushan}{\Google}

\bigskip

\xGoogle
\xCornell
\xPrinceton
\xPCTS
\xStanford
\xUCONN
\xUMass
\xAU
\xCQC
\xUCR
\xCU
\xUoCA
\xLjubljana

{\hypertarget{corra}{${}^\ddagger$} These authors contributed equally to this work.}\\

{\hypertarget{corrb}{${}^\mathsection$} Corresponding author: tomaz.prosen@fmf.uni-lj.si}\\

{\hypertarget{corrb}{${}^\mathsection$} Corresponding author: pedramr@google.com}

\end{footnotesize}
\end{flushleft}

\twocolumngrid


\newpage
\onecolumngrid




\setcounter{equation}{0}
\setcounter{figure}{0}
\setcounter{table}{0}
\setcounter{page}{1}
\makeatletter
\renewcommand{\thesection}{S\arabic{section}}
\renewcommand{\theequation}{S\arabic{equation}}
\renewcommand{\thefigure}{S\arabic{figure}}
\renewcommand{\thetable}{S\arabic{table}}

\newpage

\begin{center}
    \textbf{\large Supplementary Materials for \\``Dynamics of magnetization at infinite temperature in a Heisenberg spin chain"}
\end{center}
\vspace{18pt}

\section{Materials and methods}
\subsection{Experimental techniques and device characterization}
\subsubsection{Overview}
\label{sec:overview}
In this section, we provide an overview of the experiment performed in our work, including the observables that are studied.
The experiments are performed using 46 frequency-tunable superconducting transmon qubits. The hardware architecture is described in Ref.~\cite{supremacy2019}. The qubits are prepared in a random bitstring state according to the probabilities set by the initial imbalance $\mu$: qubits on the left side of the chain are prepared in $|1\rangle$ with probability $p=e^{\mu}/(e^\mu + e^{-\mu})$, otherwise $|0\rangle$, and qubits on the right are prepared in $|0\rangle$ with probability $p$, otherwise $|1\rangle$. The system is then evolved with alternating layers of fSim gates\,\cite{Brooks2020,Neill2021accurately,bound_states}, which implment a Floquet XXZ model. Finally, all 46 qubits are measured in the computational basis. Because ideal fSim gates are number-conserving, we post-select on the measured bitstrings having the correct number of 1s, effectively mitigating against photon loss, which otherwise causes the number of 1s to decay. After sampling over $N_{\rm states}$ initial bitstring states, we compute the expectation value of the $k$th power of the transferred magnetization as

\begin{equation}
    \label{eq:powers}
    \langle \mathcal{M}(t)^k \rangle = \frac{1}{N_{\rm states}} \sum_i \frac{1}{N^{\rm counts}_i(t)} \sum_j  N^{\rm counts}_{ij}(t) \left(2\left(N_1^R(j) - N_1^R(i) \right)\right)^k,
\end{equation}
where $i$ is the initial bitstring and $j$ is the measured bitstring, $N_i^{\rm counts}(t)$ is the total number of counts that survive postselection after $t$ cycles when the initial state is $i$, and $N^{\rm counts}_{ij}(t)$ is the number of times the bitstring $j$ is measured after $t$ cycles when the initial state is $i$. $N_i^{\rm counts}(t) = \sum_j N^{\rm counts}_{ij}(t)$. $N_1^R(i)$ is the number of 1s in the right half of the binary representation of $i$. Moments are computed as
\begin{equation}
\begin{split}
    \alpha_k(t) &= \left\langle \left(\mathcal{M}(t) - \langle \mathcal{M}(t)\rangle\right)^k\right\rangle\\
    &= \sum_{i=0}^k \begin{pmatrix} k \\ i \end{pmatrix} \left\langle \mathcal{M}(t)^{k-i} \right\rangle \left( - \left\langle \mathcal{M}(t) \right\rangle \right)^i,
\end{split}
\end{equation}
where the second line is written in terms of the experimentally measured quantities, Eq.~\eqref{eq:powers}.
Finally, the skewness $\mathcal{S}(t)$ and kurtosis $\mathcal{Q}(t)$ are computed as
\begin{equation}
\begin{split}
    \mathcal{S}(t) &= \frac{\alpha_3(t)}{ \alpha_2(t)^{3/2}} \\
    \mathcal{Q}(t) &= \frac{\alpha_4(t)} {\alpha_2(t)^2} - 3.
\end{split}
\end{equation}
Statistical uncertainties of each of these quantities are computed using the remove-one jackknife method, wherein one initial state is removed from the sample, and the variation of the quantity of interest (e.g. of $\langle \mathcal{M}(t)\rangle$, $\alpha_2(t)$, $\mathcal{S}(t)$, or $\mathcal{Q}(t)$), depending on which state is removed, is used to estimate the statistical uncertainty of that quantity. The jackknife method is also used to estimate bias, which is found to be negligible compared to the statistical uncertainties. The jackknife method is described in more detail in Section \ref{sec:jackknife}.

\begin{figure}[h]
    \centering
    \includegraphics[width=\textwidth]{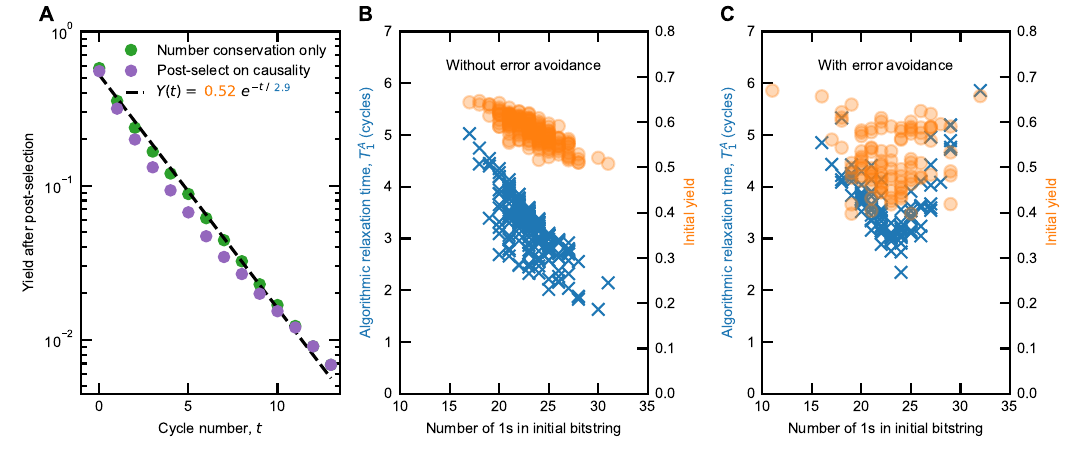}
    \caption{{\bf A.} The fraction of the total counts that pass the post-selection, as a function of cycle number, for a single initial bitstring. The green markers (higher points) indicate the yield when post-selecting only on number conservation. The purple markers (lower points) indicate the yield when post selecting on whether the observed bitstring is causally possible given the initial bitstring in the ideal circuit, a more stringent criteria than number conservation. Past cycle 11, the causality filter has no effect beyond enforcing number conservation (for this particular bitstring and set of measurements). A decaying exponential is fit to the yield when post-selected on number conservation only. We call the yield at $t=0$ the initial yield, which is less than one because of readout error. We call the decay constant the algorithmic relaxation time, $T_1^A$, which is a measure of the effective amplitude damping rate in the experiment. {\bf B.} The initial yield (orange circles) and algorithmic relaxation time (blue $\times$s) as a function of the number of 1s in the initial bitstring. As expected, the algorithmic relaxation time decreases as the number of 1s increases. The initial yield also decreases with the number of 1s in ithe initial bitstring because $0\to1$ readout error rates tend to be lower than $1\to0$ readout error rates. {\bf C.} Employing the error avoidance technique in which initial bitstrings $b_i$ with more than $N_Q/2 = 23$ 1s are replaced with $\bar b_i$ and the resulting measured bitstrings $b_f$ are replaced with $\bar b_f$, so that the number of 1s physically implemented is never more than 23. (Or equivalently, we relabel the $|1\rangle$ and $|0\rangle$ states when the number of 1s is more than 23.) As expected, this results in a V shape for $T_1^A$ as a function of the number of 1s in the initial bitstring, with the minimum at 23 1s, corresponding to half-filling. At cycle 23, the yield at half-filling is about $2\times 10^{-4}$, and we use $6\times 10^4$ shots per initial state, resulting in about 12 shots per initial state after post-selection. We average over approximately 100 initial states for each value of $\mu$, $\theta$, and $\phi$ that we study.}
    \label{fig:S1}
\end{figure}

In addition to post-selecting on number conservation, we apply several additional error avoidance and mitigation techniques:

(1) We post-select on the causal possibility of the observed bitstring. For example, if an initial bitstring on 8 qubits is 11011000, then it is not possible, in a noiseless system, to observe 01011001 at cycle 1. The rightmost 1 appeared acausally (likely by readout error) and hence the bitstring is filtered out even though it contains the correct number of 1s. We have an efficient algorithm (described in Section \ref{sec:causal_filter} for checking whether a given observed bitstring is causally possible after $t$ cycles from a given initial bitstring. This filtering mostly affects the earliest few cycles, for which the number of causally connected bitstrings is small. This filtering is a small effect compared to the number-conserving post-selection, which keeps an exponentially decaying number of bitstrings as a function of cycle number (see Figure \ref{fig:S1}A).

(2) Noting that the effects of amplitude damping ($T_1$) on our experiment are worse in initial bistring states with more 1s, when an initial state is more than half-full, i.e. the number of 1s is greater than 23, we relabel the $|0\rangle$ and $|1\rangle$ states, i.e. we start in the initial bitstring $\bar b_i$ instead of $b_i$ and then replace each measured bitstring $b_j$ with $\bar b_j$, where $\bar b$ means to apply a NOT operator to all of the bits in $b$. The advantage gained from this technique is illustrated in Figure \ref{fig:S1}C.

Figure \ref{fig:S1}A shows exponential decay of the fraction of counts that survive the post-selection. We call the decay constant the algorithmic relaxation time, $T_1^A$, which, as illustrated in Figure \ref{fig:S1}C, is about 3 cycles at half filling. If one naively estimates $T_1^A$ at half-filling from
\begin{equation}
    e^{-t/T_1^A} \overset{?}{=} \prod_{i=1}^{23} e^{-t/T_1^{(i)}} \Longrightarrow \frac{1}{T_1^A} \overset{?}{=} \sum_{i=1}^{23} \frac{1}{T_1^{(i)}},
\end{equation}
where $T_1^{(i)}$ is the $T_1$ of qubit $i$, measured at its idle frequency, we obtain an estimate for $T_1^A$ of over 7 cycles, even if we pick sum to be over the 23 qubits with the shortest $T_1$ out of the 46 total.

\begin{figure}[h]
\centering
\includegraphics[width=\textwidth]{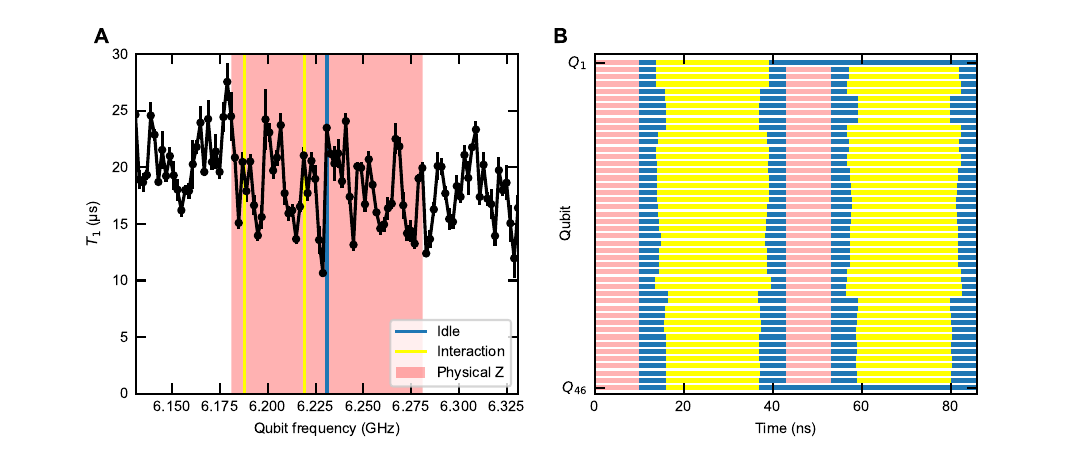}
\caption{{\bf A} $T_1$ as a function of frequency for a typical qubit in our chain of 46 qubits. Vertical lines indicate the idle frequency and the interaction frequencies with each of the two neighboring qubits. The shaded region indicates the range of frequencies used to obtain physical Z rotation angles from $-\pi$ to $\pi$. {\bf B} Schematic showing the qubit frequency as a function of time during one cycle for all 46 qubits in a typical calibration used during this experiment.}
\label{fig:S2}
\end{figure}

There are two main mechanisms expected to cause discrepancies between the algorithmic T1 and the estimate based on single-qubit T1 values. First, when the coupling is turned on, the coupler is brought close to the qubits in frequency, allowing noise in the coupler to affect the qubit. This can also enable noise-induced transitions from the qubit to the coupler. Second, the relevant $T_1$ for the experiment is not the $T_1$ at the idle frequency, even though that is what is typically optimized for and reported. The fSim gates are implemented as in Refs.~\cite{bound_states, MIP}; pairs of qubits are tuned to their interaction frequencies in a trapezoidal coupler pulse, the amplitude and duration of which are tuned to obtain the desired SWAP and controlled-phase angles. The resulting fSim gate includes single-qubit phases, which must be calibrated to zero by applying physical Z rotations\,\cite{bound_states}. Physical Z gates are fixed-duration 10-ns gates in which the qubit frequency is detuned from the idle frequency $f_0$ to the frequency $f_z$. In the frame rotating at the idle frequency, the qubit accumulates a phase of $2\pi (f_z - f_0) \times 10 \, {\rm ns}$. Therefore, the full range of phases from $-\pi$ to $\pi$ can be obtained by the range of frequencies $|f_z - f_0| \leq 0.05\, {\rm GHz}$. Figure \ref{fig:S2}A shows $T_1$ as a function of frequency for a typical qubit in our chain, indicating the idle frequency, the interaction frequencies with the two neighboring qubits, and the range of frequencies used for physical-Z rotations. It is readily seen that, although the idle frequency may be optimized to give a long $T_1$, other frequencies used during the circuit execution have $T_1$s that can be about a factor of 2 shorter. Figure \ref{fig:S2}B shows how the frequencies for each of the 46 qubits in our chain vary over the course of a cycle. Clearly, $T_1$ during the circuit execution differs from $T_1$ at the idle frequency, and the factor-of-two difference between the measured and predicted algorithmic relaxation time is plausibly explained.

\begin{figure}[h]
\centering
\includegraphics[width=0.5\textwidth]{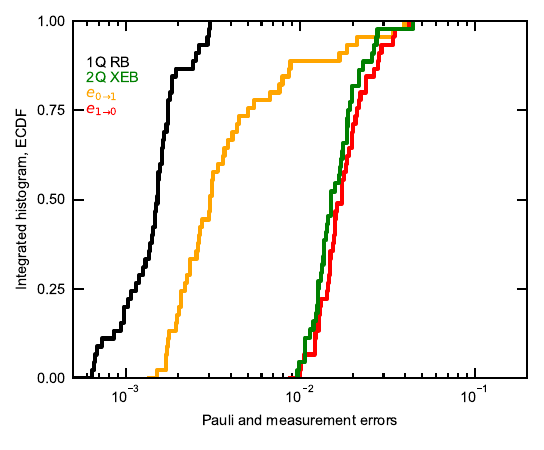}
\caption{Typical error rates achieved on our 46-qubit chain, showing readout error, single-qubit randomized benchmarking (RB), and two-qubit cross-entropy-benchmarking (XEB) cycle Pauli errors. The slightly lower error rates reported in Ref.~\cite{bound_states} are due to the different sets of angles used.}
\label{fig:S3}
\end{figure}

Post selection largely mitigates against amplitude damping, characterized by the algorithmic relaxation time, at the cost of an exponential overhead in the number of shots required. However, some errors make it past the post selection. In particular, although amplitude damping causes the number of 1s to decrease, $0\to 1$ readout errors cause it to increase. Therefore, when both amplitude damping and $0\to 1$ readout error occur, the measured bitstrings can have the correct number of 1s and pass the post-selection. As evident in Figure \ref{fig:S1}A, by later cycles, the vast majority of bitstrings have had some amplitude damping, and on 46 qubits, it is likely that at least one $0\to 1$ readout error will occur (typical readout error rates are shown in Figure \ref{fig:S3}), so this is a non-negligable effect. It manifests as excitations appearing to jump nonlocally along the chain, moving from the side with high concentration to the side with low concentration faster than they would without noise. As described in Section \ref{sec:noisy_sim}, we perform simulations including this effect and find that it explains most of the discrepancy between the noiseless simulation and the experiment. Evidently, the quality of our post-selected experimental results could be improved by changing how the readout calibration is done; the readout centers could be chosen to decrease the $0\to 1$ error rate at the expense of the $1\to 0$ error rate (and hence requiring more shots). We leave this modified calibration technique for future work.

There are other sources of errors that are not mitigated by post-selection. These include dephasing, leakage (occupation of the $|2\rangle$ state), and control errors (fSim angle miscalibrations). Because the coupling strengths used here are not particularly high, leakage is not expected to be a dominant source of error. We characterize single- and two-qubit dephasing, as well as control errors, and include these effects in our simulations. However, we find that most of the observed discrepancy is explained by $T_1$ and readout errors alone (Figure \ref{fig:S4}).

The data included in this paper were collected over the course of several months, on different sets of qubits and two different devices. In order to ensure consistent data quality, readout error rates, single-qubit error rates, and two-qubit cross-entropy benchmarking (XEB) fidelities were measured periodically, typically after every 10 initial states; we kept only data for which the maximum error rates were below thresholds that we set.

\newpage
\newpage
\subsubsection{Causal filter}
\label{sec:causal_filter}

Here we present the causal filter with which we post-select our measured bitstrings. We illustrate the algorithm with the two bitstrings mentioned in Section \ref{sec:overview}. Suppose that the initial bitstring is 11011000 and we want to know whether 01011001 is a possible measurement outcome after one cycle.

\begin{figure}[h]
    \centering
    \includegraphics{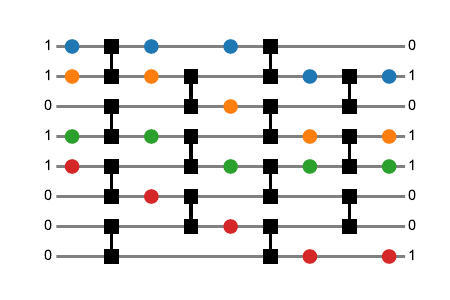}
    \caption{An illustration of how we determine the minimum number of cycles needed to obtain a given final bitstring from a given initial bitstring. In this case, at least 1.5 cycles are needed to obtain 01011001 from an initial bitstring of 11011000 under the ideal dynamics.}
    \label{fig:causal_filter}
\end{figure}

To answer this question, we determine the minimum number of cycles required to obtain 01011001 in the ideal dynamics. First, we assign identities to the ``1"s in the initial and final bitstring determined by their order. This is illustrated in Figure \ref{fig:causal_filter}, where we assign colors to the ``1"s in both the initial bitstring (on the left) and the final bitstring (on the right). We then consider each layer of fSim gates and move the excitations if it is allowed by the gates and if doing so brings the excitations closer to their desired locations. For example, in the first layer, the blue and orange excitations are blocking each other from moving (since  color is determined by order). The green one could move up, but doing so would bring it further from its final position, so it does not move. The red excitation moves down because that brings it closer to its final position. From the figure, it is clear that at least 1.5 cycles are needed to obtain 01011001 from 11011000. Therefore, this bitstring would be filtered out if seen after only one cycle.

\subsubsection{Gate calibration}
\begin{figure}[h]
\centering
\includegraphics[width=0.64\textwidth]{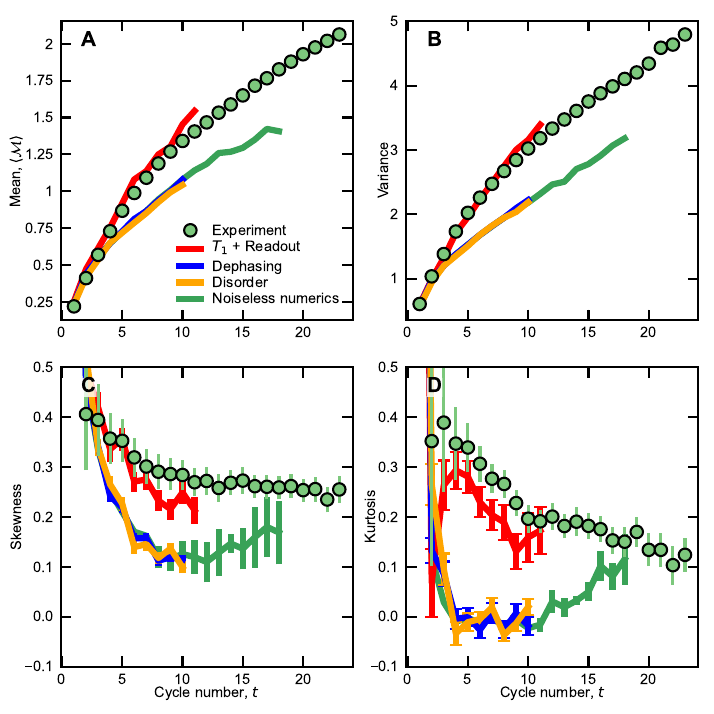}
\caption{Effects of various sources of error on the first four moments of the transferred magnetization, shown for the same parameters as in the $\Delta > 1$ case in Figure 2 of the main paper (fSim angles $(\theta, \phi) = (0.17\pi, 0.6\pi)$ and initial imbalance $\mu=0.5$). The red curve shows the effects of amplitude damping and readout error. The amplitude damping rate, $T_1$, is extracted from the experimental data as in Figure \ref{fig:S1}A and is assumed to be the same for all qubits. Readout error rates are measured by sampling random bitstrings on all 46 qubits. Dephasing is characterized using a Floquet protocol. Disorder refers to miscalibrations of the angles $\theta$ and $\phi$ as well as the single-qubit $Z$ phases (see Eq.~S2 of \cite{bound_states}). Miscalibrations of these angles are measured using unitary tomography. Error bars are estimated using the delete-one jackknife method described in Section \ref{sec:jackknife}. It is clear that amplitude damping and readout constitute the dominant source of error. This is encouraging for future experiments because the overall error rate is controlled by the $0\to1$ readout error rate, which can be decreased at the expense of the $1\to0$ error rate.}
\label{fig:S4}
\end{figure}

\begin{figure}
    \centering
    \includegraphics[width=0.32\textwidth]{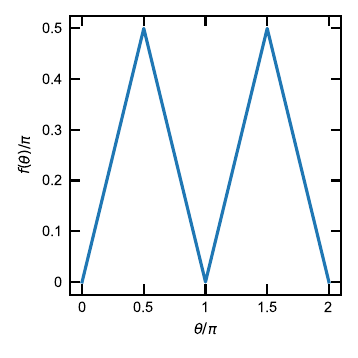}
    \caption{The periodic triangle wave function $f$ appearing in Eq.~\eqref{eq:polyfit}.}
    \label{fig:sawtooth}
\end{figure}

\begin{figure}
    \centering
    \includegraphics[width=0.85\textwidth]{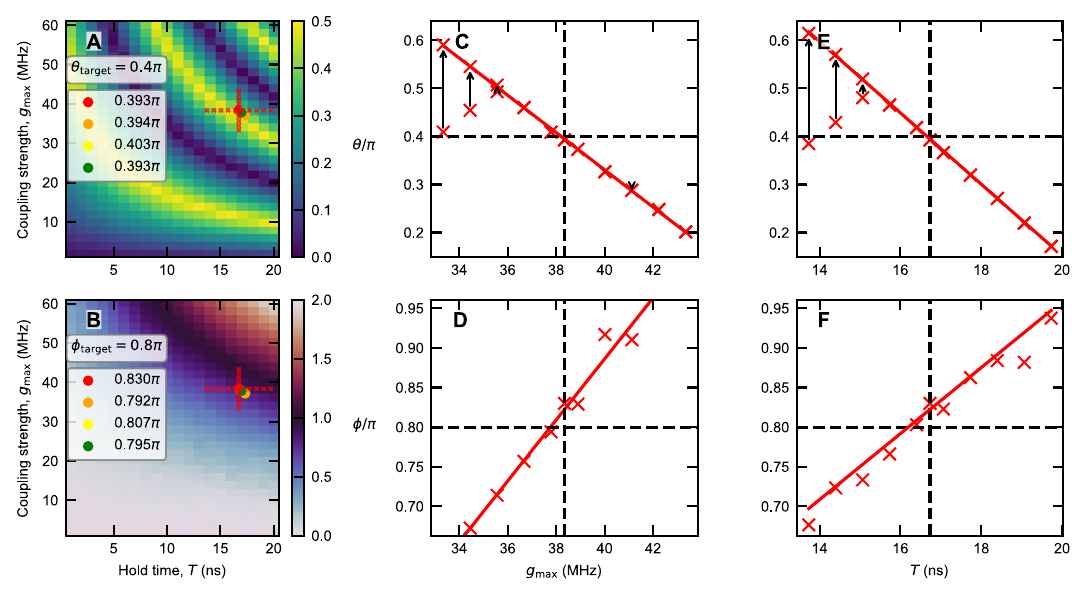}
    \caption{Illustration of the iterative calibration of fSim gates. Panels {\bf A} and {\bf B} show, in the background, the results of a 2D grid sweep, characterizing the fSim angles $\theta$ and $\phi$, respectively, as functions of the hold time, $T$, and coupling strength, $g_{\rm max}$,  of the trapezoidal pulse, for a particular qubit pair. (The rise time of the trapezoid is fixed at 5 ns.) The red colored marker indicates an initial guess for $T$ and $g_{\rm max}$. In this case, the initial guess comes from the 2D grid sweep, but an advantage of the iterative technique is that the grid sweep is not needed. More often, we provide the initial guess without ever doing a 2D sweep, allowing faster calibration of gates. The measured fSim angles for this guess are shown in the boxes on the left. We then measure the fSim angles along the cross shape indicated by the small red xs in panels {\bf A} and {\bf B}. The results of these measurements are shown in panels {\bf C}-{\bf F}, with panels {\bf C} and {\bf D} showing the constant-$T$ part of the cross, and panels {\bf E} and {\bf F} showing the constant-$g_{\rm max}$ part. For the $\theta$ fits, the triangle-wave ($f$ in Eq.~\eqref{eq:polyfit}) is unwound as indicated by the arrows. This fitting procedure gives us the six parameters in Eq.~\eqref{eq:polyfit}. We then invert the fitted polynomial to obtain the next guess for $T$ and $g_{\rm max}$, indicated by the orange point in panels {\bf A} and {\bf B}. Sometimes, no real solutions exist within the physical bounds that we place on $T$ and $g_{\max}$, in which case we find the closest approach, minimizing the cost function $C = (\phi-\phi_0)^2/\phi_0^2 + (\theta - \theta_0)^2/\theta_0^2$, where $\theta_0$ and $\phi_0$ are the target angles, subject to the bounds that we place on $T$ and $g_{\rm max}$. The cross sweep is iterated several times (in this case 3 times, resulting in 4 successive guesses), and the best guess (the one minimizing the cost function) is used.}
    \label{fig:S6}
\end{figure}

We implement fSim gates using the same trapezoidal coupler pulses used in previous works \cite{Neill2021accurately,bound_states, MIP}. Here, we note some differences between how gates are calibrated here versus in previous works. (1) in Ref.~\cite{bound_states}, the fSim gates were not of uniform duration across pairs. Here, we adjust the padding (the idle time in Figure \ref{fig:S2}B) so that the gate duration (including the added padding) is uniform across pairs, leading to the neat alignment in time acros qubits shown in Figure \ref{fig:S2}B. (2) In Ref.~\cite{bound_states}, Floquet calibration of the fSim angles allowed them to be controlled with high precision. Because control errors (called disorder in Figure \ref{fig:S4}) are a negligible source of error for us, we instead use unitary tomography to calibrate the fSim angles. This enables us to calibrate gates quickly and in a way that is mostly automated. We tried Floquet calibration, which allows for more precise calibrations of the angles at the cost of a higher overhead, but found that it did not improve our gate fidelities. (3) We iteratively calibrate the hold time $T$ and coupling strength $g_{\rm max}$ of the trapezoidal pulse by measuring the fSim angles $\theta$ and $\phi$ at points in a small cross shape in the $(T, g_{\rm max})$ plane centered at the previous guess, fitting the polynomials
\begin{equation}
\label{eq:polyfit}
\begin{split}
    \theta &= f((b_1 g_{\rm max} + b_0)(T + T_b))\\
    \phi &= (c_1 g_{\rm max} + c_0)(T + T_c),
\end{split}
\end{equation}
where $f$ is the triangle-wave function illustrated in Figure \ref{fig:sawtooth}. This technique allows for fast calibration of fSim gates without relying on expensive 2d sweeps and in a way that is more robust to noise than gradient descent. Figure \ref{fig:S6} illustrates this calibration procedure.

An advantage of being able to quickly optimize $T$ and $g_{\rm max}$ for the desired fSim angles is that we can now put this in an outer loop that adjusts the interaction frequencies. Indeed, during gate calibration, we set a minimum two-qubit XEB fidelity and optimize the interaction frequencies (re-optimizing $T$ and $g_{\rm max}$ each time) until all qubit pairs achieve the desired fidelity.

\newpage

\subsubsection{Jackknife estimate of uncertainties}
\label{sec:jackknife}
We use the ``delete one" jackknife to estimate the statistical uncertainty of our skewness and kurtosis, and dynamical exponent measurements \cite{Quenouille1949, Tukey1958, JackknifeReview}. For finite $\mu$, where we average over initial states, define $\hat \theta_{(i)}$ to be the quantity of interest, for example the skewness or kurtosis, computed with initial state $i$ removed. Define $\hat \theta_{(.)} = \frac{1}{N_s} \sum_i \hat \theta_{(i)}$, where $N_s$ is the number of initial states. Then we use
\begin{equation}
\label{eq:jackknife}
\sigma_{\hat \theta} = \sqrt{\frac{N_s - 1}{N_s} \sum_{i} \left( \hat \theta_{(i)} - \hat \theta_{(.)}\right)^2}
\end{equation}
as our estimate of the statistical uncertainty of the quantity $\hat \theta$.

The averages of skewness or skewness over cycle number are uncertainty-weighted averages, and the uncertainty of the average is computed directly using Eq.~\eqref{eq:jackknife}.

\begin{figure}[h]
    \centering
    \includegraphics{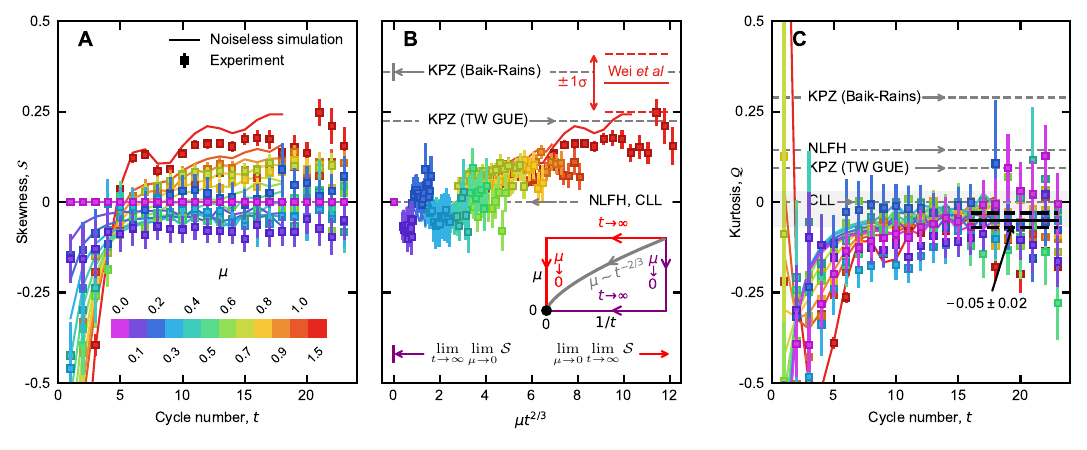}
    \caption{Same as Figure 3 but showing the statistical uncertainty of each data point ($1\sigma$ error bars). Note that, because the same initial bitstrings are used across cycles, the errors at different cycles for the same value of $\mu$ are not independent.}
    \label{fig:error_bars}
\end{figure}

In the case of $\mu=\infty$, there is only one initial state, so we instead perform the jackknife estimate by deleting each one of the shots. This gives us the uncertainty of the skewness and kurtosis at each cycle. Because the shots, unlike the initial states, are different at each cycle number, we cannot use the jackknife to directly compute the uncertainty of the cycle-averaged skewness and kurtosis for $\mu=\infty$. Instead, in this case, we treat the skewness and kurtosis at each cycle as independent random variables, so the uncertainty of their weighted average is
\begin{equation}
\sigma_{\rm weighted\, avg} = \frac{1}{\sqrt{\sum_t w_t}},
\end{equation}
where the weight, $w_t$, at cycle $t$ is $1/\sigma_t^2$, where $\sigma_t$ is the uncertainty of the quantity (either skewness or kurtosis) at cycle $t$. Figure \ref{fig:error_bars} shows the experimental data from Figure 3, including statistical error bars.

\newpage

\subsection{Simulation techniques and numerical results}
Our main quantity of interest, the transferred magnetization $\mathcal{M}$, counts the number of excitations that have moved across the center of the 1D chain. A significant simplification stems from the fact that, at early times, excitations far from the center have not had time to cross the center of the chain. As a result, we can imagine an infinitely large system and only simulate a finite number of sites in order to study it. In particular, at cycle $t$, it is only necessary to simulate $2t$ sites \cite{Note1}. As a result, the optimal simulation technique varies depending on the cycle number. Through cycle 8 (16 qubits), we obtain exact results by simulating the full density matrix. Beyond that point, density matrix simulations become costly, so for cycles 9--18 (18--36 qubits), we instead sample random initial bitstring states, as done in the experiment, and apply exact statevector simulation to these initial states. For the pure domain wall case ($\mu=\infty$), we employ tensor-network simulations using the time-evolving block decimation (TEBD) algorithm \cite{TEBD} to extend the simulations to cycle 23.

 \subsubsection{Analytical results}
 In this section, we derive certain analytical results for small depths.
 At very early times, the matrices involved are small enough that it is possible to obtain relatively simple analytical results. Some of these are tabulated here:
 \paragraph*{Cycle 1}
 To compute the transferred magnetization at cycle 1, it is only necessary to consider two qubits with an fSim gate between them. Therefore, the probability distribution of the transferred magnetization takes a simple form. For positive integer power $k$, we have:
 \begin{equation}
     \langle \mathcal{ M}^k \rangle = \begin{cases} 2^k \sin^2 \theta \tanh \mu, & k\,\, \mbox{odd}\\
      2^{k-1} \sin^2\theta \left(1 + \tanh^2 \mu\right), & k\,\, \mbox{even} \end{cases}.
 \end{equation}
In particular, the mean, variance, and skewness are
\begin{equation}
\label{eq:cycle1}
\begin{split}
    \langle \mathcal{ M} \rangle &= 2 \sin^2 \theta \tanh \mu \\
    \mbox{var} &= 2 \sin^{2}\theta \left(1 +\cos(2\theta)\tanh^{2}\mu  \right)\\
    \mathcal{S} &= \frac{2 \sqrt{2} \left(\left(2 \sin^{4}\theta \tanh^{2}\mu + 1\right) \left(\sinh{\left(2\mu \right)} + \cosh{\left(2 \mu \right)} + 1\right)^{2} - 3 \left(\sinh{\left(4 \mu \right)} + \cosh{\left(4 \mu \right)} + 1\right) \sin^{2}{\left(\theta \right)}\right) \tanh \mu}{\left(\sinh{\left(2 \mu \right)} + \cosh{\left(2\mu \right)} + 1\right)^{2} \sqrt{\left(\cos{\left(2 \theta \right)} \tanh^{2}\mu + 1\right)^{3}} \sin \theta}\\
    &= \mu\sqrt{2} (2\csc \theta -3\sin\theta)  + O(\mu^3)\\
    \mathcal{Q} &= 2\csc^2\theta - 3 + O(\mu^2)
\end{split}
\end{equation}
Observe that, for $\mu \ll 1$, the skewness and kurtosis are both positive for small $\theta$, i.e. the Trotter limit, and negative for large $\theta$. The crossover happens at $\theta = \arcsin(\sqrt{2/3}) \approx 0.3 \pi$. In Figure 3 of the main text, we choose $\theta = 0.4\pi$, which is why we observe negative skewness and kurtosis. In continuous-time Hamiltonian dynamics, we expect the opposite signs.

\paragraph*{Cycle 2}
It is also possible to obtain analycial expressions at cycle 2. In particular, the mean and variance of the transferred magnetization are
\begin{equation}
\begin{split}
 \langle \mathcal{ M} \rangle &= 2\mu \sin^{2}\theta\left(\cos^4 \theta (3 + \cos \phi) + 2 \sin^2 \theta\right)  + O\left(\mu^{3}\right)\\
 \mbox{var} &= \sin^4 \theta (1-\cos\phi) + \frac{1}{8} (3+\cos\phi)(7 \sin^2 \theta + \sin^2(3\theta)) + O(\mu^2)
\end{split}
\end{equation}
 
\subsubsection{Simulation cost and runtime}
\begin{figure}[h]
    \centering
    \includegraphics{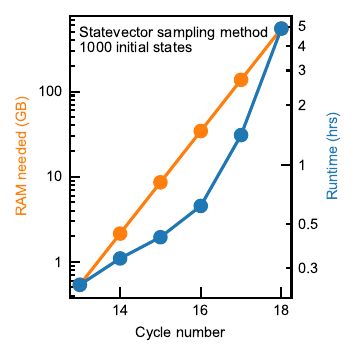}
    \caption{The resources needed to generate 1000 initial states using our implementation of the statevector sampling method on up to eight 80-GB NVIDIA A100 GPUs using NVIDIA cuQuantum \cite{cuQuantum}.}
    \label{fig:resource_estimate}
\end{figure}

To perform statevector simulations out to cycle 18 (36 qubits), we use NVIDIA's cuQuantum \cite{cuQuantum} and its interface with qsim \cite{qsim}. cuQuantum supports multi-GPU quantum simulations, and with eight 80-GB NVIDIA A100 GPUs, available to virtual machines running in Google Cloud's compute services, we can simulate up to 36 qubits. On this platform, a noiseless 18-cycle simulation takes about 17.6 seconds per initial state. However, the memory required to store the state increases exponentially in the cycle number, as shown in Figure \ref{fig:resource_estimate}. The cuQuantum implementation stores $2^{2t}$ complex numbers. The memory footprint could be reduced by taking advantage of number conservation, in which case only $\begin{pmatrix} 2t \\ t\end{pmatrix}$ complex numbers would be needed to represent the state.

\subsubsection{Noisy simulations}
\label{sec:noisy_sim}
In this subsection, we describe how we performed the noisy simulations shown in Figure \ref{fig:S4} and Figure 2 of the main text. We note that these are classical simulations that include the effects of noise, not our experiment. The simulations with disorder were performed in a straightforward way; we simply measured the actual fSim angles, including the single-qubit phases, using unitary tomography, and used the measured angles in the simulation, simulating only the $2t$ qubits about the center, as in the noiseless statevector simulations. The simulations with dephasing were also performed in a straightforward way; we simply averaged over many circuits, adding high- and low-frequency Gaussian noise to the fSim angles, as well as Z-rotations between the fSim gates with random angles that vary both within a circuit and across shots.

The simulation of amplitude damping and readout error is slightly more involved. For each initial 46-qubit bitstring, we consider separately the $2t$ qubits about the center and the remaining $46-2t$ outer qubits. For the center qubits, we perform a noisy simulation using cirq/qsim \cite{qsim} that includes the measured amplitude damping as gates applied between the layers of fSim gates. The outer qubits are treated as if no two-qubit gates are applied; qubits prepared in $|1\rangle$ are stochastically flipped to $|0\rangle$ with a probability $1-e^{-t/T_1}$. The resulting bitstrings from the center and outer qubits are concatenated back together, and then bits are randomly flipped according to the $0\to1$ readout error rate $e_0$ and the $1\to0$ readout error rate $e_1$. Finally, the same post-selection that is used in the experiment is applied to the simulated bitstrings, so that only those conserving the number of 1s and satisfying the causality constraints survive. The readout error rates used here are those measured on the device at the time the experiment was run, including the qubit-by-qubit variations. The amplitude damping rate, $T_1$, is obtained as in Figure \ref{fig:S1}A and is approximated as being the same across all qubits.

\subsubsection{Length independence}
As demonstrated in Table \ref{tab:tableS1}, the transferred magnetization is independent of the length of the chain as long as the chain consists of at least $2t$ qubits, where $t$ is the cycle number.

\definecolor{Gray}{gray}{0.9}

{
\setlength{\tabcolsep}{0mm}
\begin{table}[h!]\centering
\scriptsize
\begin{tabular}{c<{\hspace{2mm}}l<{\hspace{2mm}}c<{\hspace{2mm}}c<{\hspace{2mm}}c<{\hspace{2mm}}c<{\hspace{2mm}}c<{\hspace{2mm}}c<{\hspace{2mm}}c<{\hspace{2mm}}c<{\hspace{2mm}}c<{\hspace{2mm}}}\toprule
& &\multicolumn{8}{c}{Absolute difference from reference kurtosis when computed with $N_Q$ qubits.} \\
Cycle &Reference kurtosis &$N_Q=2$ &$N_Q=4$ &$N_Q=6$ &$N_Q=8$ &$N_Q=10$ &$N_Q=12$ &$N_Q=14$ & $N_Q=16$ \\\midrule
1 & $-0.7888543819998315$ & 0 & $10^{-15}$ & $10^{-15}$ & $10^{-15}$ & $10^{-15}$ & $10^{-14}$ & $10^{-14}$ & NC\\ 
\rowcolor{Gray}
2 & $-0.3236513411118609$ &  & 0 & $10^{-15}$ & 0 & $10^{-15}$ & $10^{-15}$ & $10^{-15}$ & NC\\ 
3 & $-0.19032877952363814$ &  &  & 0 & 0 & $10^{-15}$ & $10^{-15}$ & $10^{-15}$ & NC\\ 
\rowcolor{Gray}
4 & $-0.13634880999637522$ &  &  &  & 0 & $10^{-15}$ & $10^{-15}$ & 0 & NC\\ 
5 & $-0.11064712866845028$ &  &  &  &  & 0 & $10^{-15}$ & 0 & NC\\ 
\rowcolor{Gray}
6 & $-0.0964617033065931$ &  &  &  &  &  & 0 & $10^{-15}$ & NC\\ 
7 & $-0.08726032489412416$ &  &  &  &  &  &  & 0 & NC\\ 
\rowcolor{Gray}
8 & $-0.08047534387997635$ & & & & & & & & 0\\
\bottomrule
\end{tabular}
\caption{Illustration of the length-independence of the transferred magnetization, as long as $N_Q\geq 2t$. Deviations are consistent with double-precision floating-point arithmetic. The values are computed using exact density matrix simulations for $(\theta, \phi) = (0.4\pi, 0.8\pi)$ and $\mu=0$, but we observe similar agreement across system sizes for all choices of these parameters and for all of the moments. Here ``NC" means ``not computed." The kurtosis at $t=1, N_Q=2$, is given by Eq.~\eqref{eq:cycle1}: $\mathcal{Q} = 2\csc^2\theta - 3$. }\label{tab:tableS1}
\end{table}
}

\newpage
\subsubsection{Data collapse}
\begin{figure}[h!]
    \centering
    \includegraphics{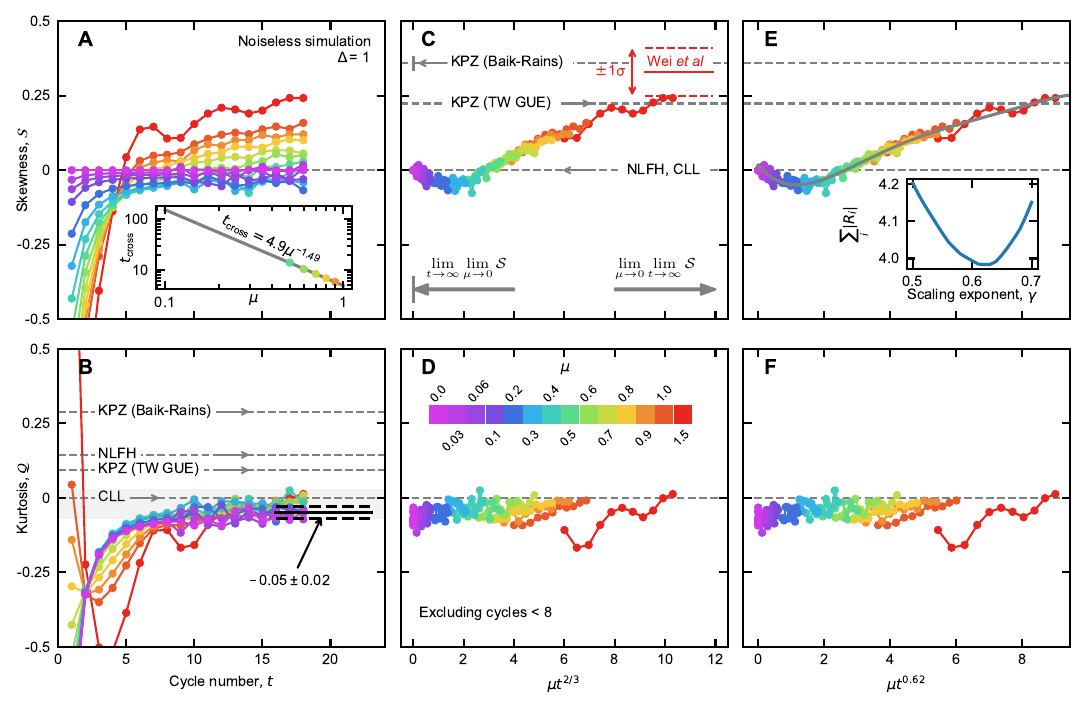}
    \caption{Data collapse. Panels {\bf A}--{\bf C} are the same as Figure 3 of the main text, but showing only data from noiseless numerical simulations. The added inset in panel {\bf A} shows the time at which the skewness becomes positive as a function of $\mu$, with a fitted power law. The fact that the fitted exponent is close to $-3/2$ supports the scaling in panels {\bf C-D}. In {\bf E}, we further explore the optimal scaling by fitting polynomials to the skewness as a function of $\mu t^\gamma$, shown for $\gamma = 0.62$ and a degree-5 polynomial. For each value of $\gamma$, we record the uncertainty-weighted sum of the residuals; i.e., the deviation of the skewness from the polynomial fit. This weighted sum of the residuals is shown versus $\gamma$ in the inset, where it is seen that $\gamma \approx 0.62$ provides the optimal data collapse. Panel {\bf F} shows the kurtosis with the same scaled x-axis. The kurtosis does not collapse, as expected; the kurtosis cannot be a function of $\mu t^\gamma$ for any positive $\gamma$ because, unlike the skewness, it has time dependence even when $\mu=0$.}
    \label{fig:Fig3_numerics}
\end{figure}

In Figure 3B of the main text, we plot the skewness as a function of $\mu t^{2/3}$. The data collapse observed there is seen even more clearly in the numerics, shown in Figure \ref{fig:Fig3_numerics}. In the inset to panel A, we see that the time at which the skewness becomes positive scales like $\mu^{-3/2}$, and, in panel B, we see a collapse of the numerical data when plotted as a function of $\mu t^{2/3}$. The power law scaling sensibly predicts that the crossing time becomes infinite as $\mu\to 0$, which makes sense because the skewness is always 0 at $\mu=0$. The kurtosis, however, cannot be a function of $\mu t^{2/3}$ because it is not constant when $\mu=0$. 

Figure \ref{fig:Fig3_numerics} shows that the skewness also appears to collapse reasonably well when plotted as a function of $\mu t^{1/3}$, thus conveying the difficulty of estimating the exponent of $t$ by eye. In order to do so in an unbiased manner, we define a quantitative measure of data collapse based on the appropriately normalized sum of fit residuals (where the fit describes the purportedly universal scaling function). Using $\mu t^\gamma$ as the scaling variable, for varying $\gamma$, we find (Fig.~\ref{fig:Fig3_numerics}) that this metric is minimized for $\gamma\sim 0.62$. This is consistent with our initial observation of $\mu t^{2/3}$ seemingly yielding the best data collapse. 
Although KPZ predicts $\gamma = 1/3$, that prediction is only valid for much smaller values of $\mu$ than studied in this experiment \cite{gopalakrishnan2019anomalous}, so it is perhaps more surprising that the data collapse at all; this is a nontrivial finding of our experiment and numerics.

\clearpage
\subsubsection{Sweeps of anisotropy and imbalance}
\begin{figure}[htb]
    \centering
\includegraphics{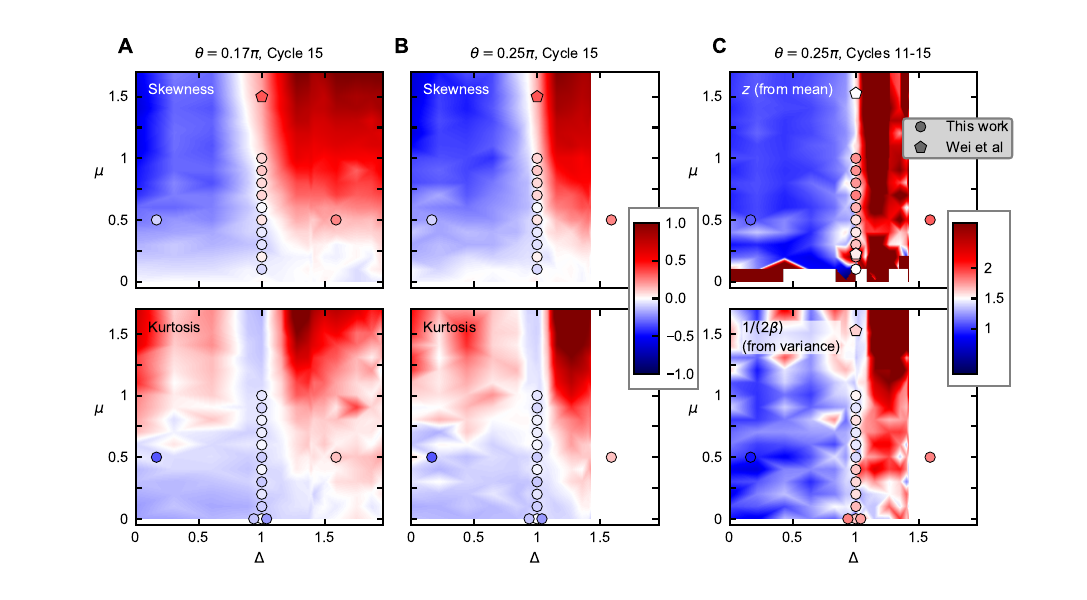}
    \caption{Skewness, kurtosis, and dynamical exponents versus initial imbalance $\mu$ and anisotropy $\Delta$. The anisotropy is a function of the two fSim angles, $\theta$ and $\phi$ (see Fig.~1). Our experimental data (circles) primarily use $\theta = 0.4\pi$ in order to achieve fast dynamics, but smaller values of $\theta$ are needed in order to achieve large $\Delta$. Therefore, in this figure, the background heatmaps show numerical results for $\theta = 0.17\pi$ (panel {\bf A}) and $\theta = 0.25\pi$ (panels {\bf B}, {\bf C}). The heatmaps are computed using 500 random initial states and interpolated between the gridded points. The experimental values of skewness and kurtosis are averaged over cycles 15 through 23, and the experimental dynamical exponents are fit to cycles 11 through 23. The experimental data point at $(\Delta, \mu) = (1.59, 0.5)$ is an exception in that it uses $\theta=0.17\pi$).}
    \label{fig:2d_sweep_heatmaps}
\end{figure}

The transport characteristics of the XXZ model are strongly dependent on the anisotropy parameter, $\Delta$. In particular, ballistic, superdiffusive and diffusive behaviors are expected in the regimes where $\Delta<1$, $\Delta=1$ and $\Delta>1$, respectively. Moreover, the KPZ conjecture was only proposed for the isotropic point ($\Delta=1$). In order to get a better sense of the parameter space in which we are operating, we performed 2D numerical sweeps of the anisotropy $\Delta$ and the initial imbalance 
$\mu$. The results are shown in Figure \ref{fig:2d_sweep_heatmaps}. They illustrate a sign change in the skewness and kurtosis close to the Heisenberg point, $\Delta=1$, as well as a clear change in the dynamical exponent.

\newpage
\section{Further experimental data}
\subsection{Dynamical exponent}
\begin{figure}[h!]
    \centering
    \includegraphics{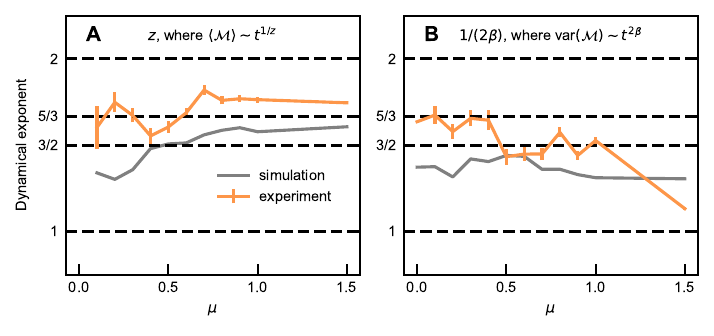}
    \caption{The dynamical exponent ({\bf A} of the mean, and {\bf B} of the variance) as a function of the initial imbalance. The KPZ prediction in both cases is 3/2. The dynamical exponents are extracted from fits over cycles 11-23 for the experiment and 11-18 for the simulation. }
    \label{fig:S7}
\end{figure}

In Fig.\,\ref{fig:S7}, we plot both the experimentally observed and numerically simulated dynamical exponents of the mean and the variance, as a function of the initial imbalance, $\mu$. At large $\mu$, we find that the dynamical exponent is higher than the superdiffusive value of 3/2. The observed values are consistent with Ref.~\cite{ljubotina2017spin}, where it was found that the dynamical exponent drifts from 3/2 at small initial imbalance $\mu$ to approximately 5/3 when $\mu$ is about 1.

\subsection{Pure domain walls}
\begin{figure}[h]
    \centering
    \includegraphics{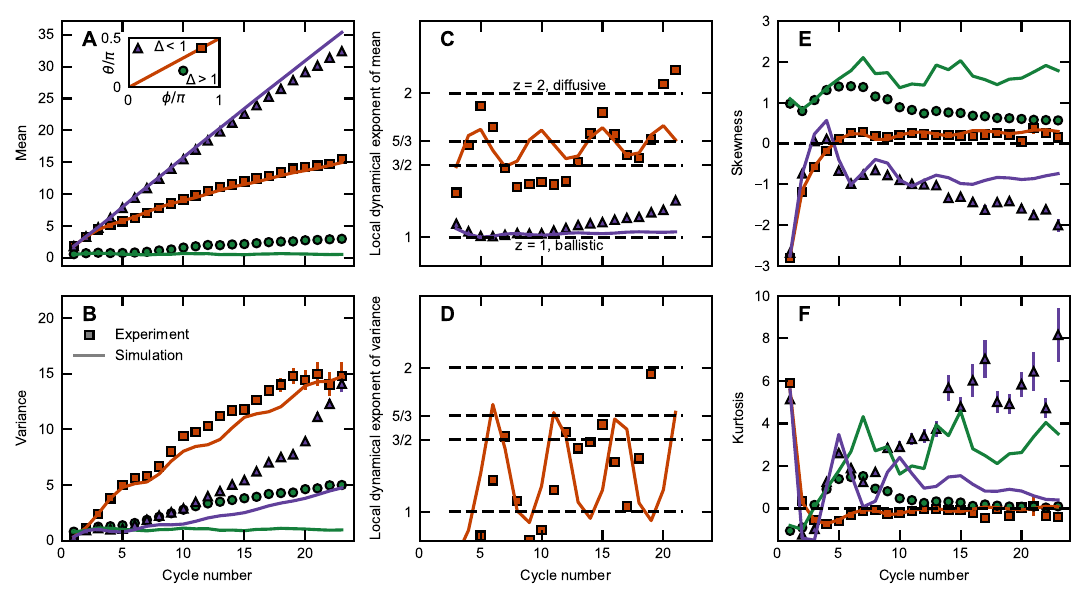}
    \caption{Experiment and simulation results for the pure domain wall initial state, $\mu=\infty$. The three pairs of fSim angles shown are $(\theta, \phi) = (0.4\pi, 0.8\pi)$ (orange squares), $(\theta, \phi) = (0.4\pi, 0.1\pi)$ (purple triangles), and $(\theta, \phi) = (0.17\pi, 0.6\pi)$ (green circles). The simulation results for cycles 19-23 are computed using TEBD.}
    \label{fig:muinf}
\end{figure}

\begin{figure}[h]
    \centering
    \includegraphics{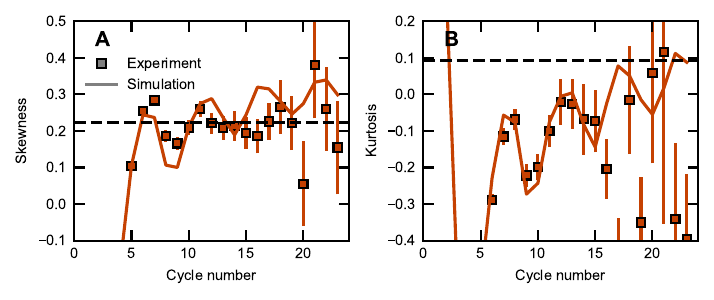}
    \caption{Experiment and simulation results for the pure domain wall initial state, $\mu=\infty$ with $(\theta, \phi) = (0.4\pi, 0.8\pi)$, i.e.~$\Delta = 1$. Horizontal dashed lines indicate the KPZ prediction for a wedge-shaped initial state (see Table \ref{tab:moments}). We show these values for comparison only. The fact that the dynamical exponent is larger than 3/2 (see Fig.~\ref{fig:muinf}C) already indicates the lack of KPZ scaling. Further, the skewness continues to increase above the KPZ prediction.}
    \label{fig:muinf_zoomed}
\end{figure}

\begin{figure}
    \centering
    \includegraphics{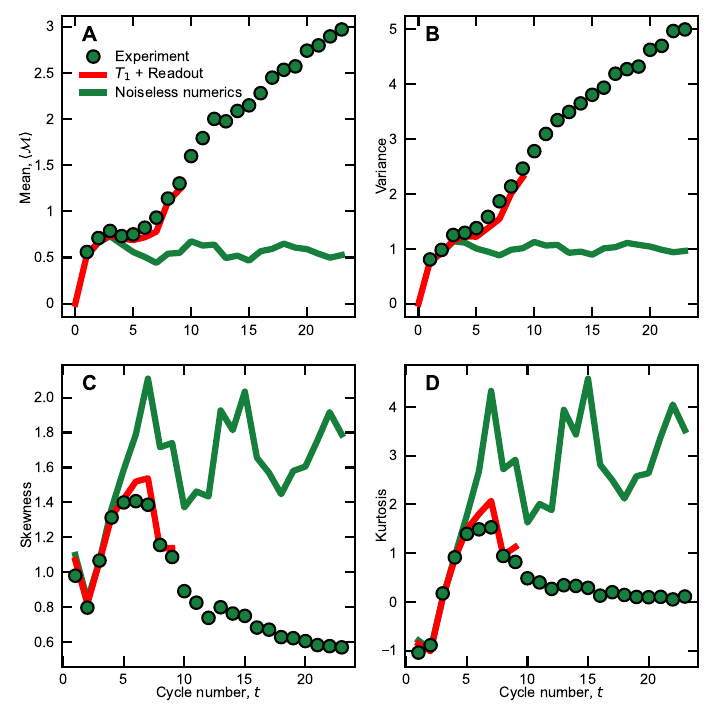}
    \caption{Comparison of experimental results for the pure domain wall simulation ($\mu=\infty$), $(\theta, \phi) = (0.17\pi, 0.6\pi)$, $\Delta = 1.6$, with noiseless and noisy simulations. The noisy simulations include only the effects of amplitude damping ($T_1$) and readout error, with $T_1$ extracted from the measured algorithmic relaxation time and the readout error rates benchmarked on the device. Compare with Figure \ref{fig:S4}, which shows similar results for $\mu=0.5$.}
    \label{fig:noisy_sim_muinf}
\end{figure}

In addition to the results presented in the main paper, we also studied the pure domain wall, $\mu=\infty$. This is a simpler experiment because it does not require any averaging over initial states. It is also easier for classical simulations; TEBD simulations converge at least to cycle 23, allowing us to check our experimental results at later cycles than is possible at finite $\mu$ with the simulation techniques used here. Experiment and simulation results are shown in Figure \ref{fig:muinf}. Our findings are largely consistent with expectations (see Section 6.1 of \cite{Bulchandani_StatMech_2021} for a review). We observe an absence of transport in the easy-axis regime ($\Delta > 1$), with the observed transport in the experiment consistent with amplitude damping and readout errors (see Figure \ref{fig:noisy_sim_muinf}). In the easy-plane regime $(\Delta < 1)$, we observe ballistic transport. In the isotropic ($\Delta = 1$) case, we observe transport with a dynamical exponent of about 5/3, consistent with the finding of Ref.~\cite{ljubotina2017spin} and differing from the 3/2 dynamical exponent of KPZ. Figure \ref{fig:muinf_zoomed} compares the skewness and kurtosis at the isotropic point with the KPZ predictions. While the skewness is close to the KPZ prediction, it can be seen in the simulation results that it continues to increase above the KPZ value. The kurtosis approaches the KPZ prediction, but it is not clear that it has stopped increasing.

\begin{figure}
    \centering
    \includegraphics{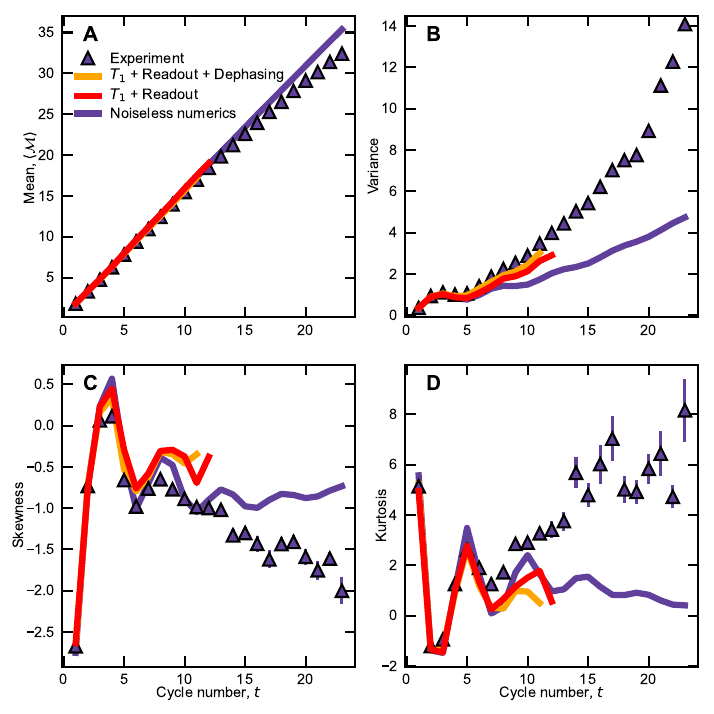}
    \caption{Sources of noise in the $\Delta = 0.16$, $\mu=\infty$ experiment. Similar to Figure \ref{fig:noisy_sim_muinf} but for $(\theta, \phi) = (0.4\pi, 0.1\pi)$. In this case, dephasing does have a noticeable effect on the variance.}
    \label{fig:noisy_sim_muinf_Dl1}
\end{figure}

\clearpage

\section{The Kardar-Parisi-Zhang universality class}
\label{sec:KPZ}
We here provide a review of the KPZ universality class, including a discussion of its limitations in describing the Heisenberg spin dynamics. 
In 1985, Mehran Kardar, Giorgio Parisi, and Yi-Cheng Zhang \cite{KPZ} set out to study the general properties of stochastically growing interfaces, examples of which include flame fronts and tumors. They abstracted these situations to a height function $h(\vec x,t)$ that obeys
\begin{equation}
    \label{eq:KPZ}
    \frac{\partial h}{\partial t} = \nu \nabla^2 h + \frac{\lambda}{2} \left(\nabla h\right)^2 + \eta(\vec x, t),
\end{equation}
where $\eta(\vec x, t)$, at each position and time, is an independent zero-mean Gaussian random variable with variance proportional to a parameter $D$. This is the stochastic Burgers equation for the slope $\nabla h$. The $\nabla^2 h$ is a diffusive term, and the $(\nabla h)^2$ is a nonlinearity. In general, an equation describing a growing interface may have higher powers of the slope, such as $(\nabla h)^4$, but these are irrelevant to the large-scale physics in the renormalization group (RG) sense. Indeed, in 1+1 dimensions ($d=1$), the KPZ equation [Eq.~\eqref{eq:KPZ}] has divergences that must be regularized, e.g., by putting it on a spatial lattice or imposing a maximum cutoff wavenumber $\Lambda$ in Fourier space. Kardar Parisi, and Zhang considered an RG flow in which (1) high-wavenumber modes are integrated out ($\Lambda \to e^{-l}\Lambda$), (2) space and time are rescaled so that the new smallest length scale is labeled by the same numerical value as the old smallest length scale ($\vec x \to e^{-l}\vec x$, $t\to e^{-zl}t$), and (3) the height function is rescaled ($h\to e^{-(d+\chi)l}h$). This procedure can be thought of as coarse graining and zooming out, while changing units so that it appears as though the zoom-out did not occur. Kardar, Parisi and Zhang showed that Eq.~\eqref{eq:KPZ} is a fixed-point of this RG flow, for specific choices of the scaling exponents and parameters. In 1+1 dimensions in particular, if $z=3/2$, $\chi=1/2$, and $\lambda^2 D/\nu^3 = 2$, then Eq.~\eqref{eq:KPZ} is invariant under the rescaling and coarse graining procedure. Further, this fixed point is stable; if $\lambda^2 D/\nu^3 > 2$, then it will flow down to 2 under the coarse graining. Conversely, if $\lambda^2 D/\nu^3 < 2$, then it will flow up to 2.

Kardar, Parisi, and Zhang therefore proposed a new universality class (the KPZ universality class) into which their equation falls. They conjectured that a variety of other systems, such as ballistic deposition (e.g., snow falling and sticking together), and the Eden model (a random growth model), are in this universality class, based on scaling exponents observed in earlier numerical experiments \cite{PlishkeRacz85, Family_1985}.

\begin{figure}[h]
    \centering
    \includegraphics{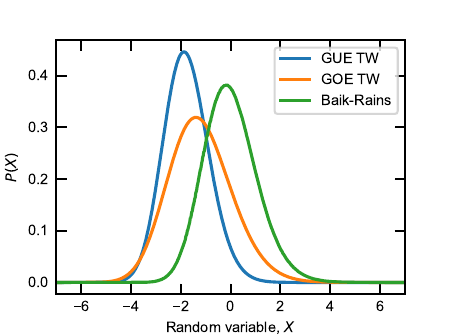}
    \caption{The Tracy-Widom (TW) and Baik-Rains probability distributions. The TW probability densities are computed using \cite{tracywidom, bejan2005largest, TW_expansion}, and the Baik-Rains probabilitiy densities are computed using those results and Ref.~\cite{BaikRains}.}
    \label{fig:TW}
\end{figure}

The KPZ equation and universality class have since been studied in great detail. For example, Kurt Johansson \cite{Johansson_2000} studied the asymmetric exclusion process, a model in this universality class, for an initial state corresponding to a wedge-shaped $h(x,t=0)$, and found that a quantity corresponding to the regularized height function, $2h(0,t) - h(\infty,t) - h(-\infty, t)$, follows the  Tracy-Widom (TW) distribution for the largest eigenvalue in the Gaussian Unitary Ensemble (GUE) at late times. Pr{\"a}hofer and Spohn \cite{PrahoferSpohn2000} generalized this result, using a mapping from the polynuclear growth model, which is in the KPZ universality class, to random permutations and from there to random Gaussian matrices to again identify the asymptotic probability distribution of the regularized height function. The precise distribution depends on the initial conditions, and there are three cases that are relevant for us: (1) \textit{flat}, meaning $h(x,0) = 0$, (2) \textit{stationary}, meaning that $h(x_i, 0) = h(x_i-1, 0) + \eta_i$, where $\eta_i$ is randomly $\pm 1$ with equal probabilities, and (3) \textit{wedge-shaped}, meaning that $h(x,0) = -|x|/\delta$. The wedge-shaped initial condition leads to the same distribution as the curved initial condition studied by Pr{\"a}hofer and Spohn \cite{Hartmann_2018} and is unaffected by nonzero variance in $h(x_i, 0)$, unlike the flat initial condition, which becomes the stationary initial condition with the addition of fluctuations. Pr{\"a}hofer and Spohn found that the asymptotic probability distributions of the regularized height function in these three cases are (1) GOE Tracy-Widom, (2) Baik-Rains, and (3) GUE Tracy-Widom, in agreement with \cite{Johansson_2000}. The corresponding values of skewness and kurtosis are listed in Table~\ref{tab:moments}. The probability distributions are plotted in Figure \ref{fig:TW}.

\begin{table}[!htp]\centering
\scriptsize
\begin{tabular}{c<{\hspace{2mm}}c<{\hspace{2mm}}c<{\hspace{2mm}}c<{\hspace{2mm}}c<{\hspace{2mm}}c}\toprule
Spin chain initial state &KPZ initial condition &Asymptotic probability distribution &Skewness &Kurtosis \\\midrule
&Flat &GOE TW &0.294 &0.165 \\
$\mu = 0$ &Stationary &Baik-Rains &0.359 &0.289 \\
$\mu > 0$ &Wedge &GUE TW &0.224 &0.093 \\
\bottomrule
\end{tabular}
\caption{Asymptotic probability distribution of the regularized KPZ height function \cite{PrahoferSpohn2000}, for the three initial conditions described in the text. The conjecture that the Heisenberg spin chain is in the KPZ universality class implies that the asymptotic transferred magnetization has the same distribution as the regularized KPZ height function. Therefore, the conjecture predicts an asymptotic skewness of 0.359 and an asymptotic kurtosis of 0.289 for the $\mu=0$ state. Note that although we included $\mu>0$ in the first column, the infinite-time dynamics of the Heisenberg chain in this case have been shown to be diffusive rather than KPZ \cite{Krajnik_PRL_2022, Gopalakrishnan_2023}.}
\label{tab:moments}
\end{table}

Pr{\"a}hofer and Spohn \cite{Prahofer2004} further used the polynuclear growth model to solve for the two point correlation correlation function,
\begin{equation}
    C(x,t) = \left\langle \left(h(x,t) - h(0,0) - t\langle\partial_t h\rangle\right)^2 \right\rangle,
\end{equation}
assuming stationary initial conditions, which imply that $C(x,0) = \frac{D}{\nu} |x|$. The slope-slope correlation function can then be obtained as
\begin{equation}
    \label{eq:KPZ_correlation_function}
   \left \langle \partial_x h(0,0) \partial_x h(x,t)\right\rangle = \frac{1}{2}\partial^2_xC(x,t).
\end{equation}
$C(x,t)$ takes the form
\begin{equation}
    C(x,t) = t^{2/3}g\left({\rm const} \cdot x/t^{2/3} \right),
\end{equation}
where $g(y)$ is a universal scaling function. Defining the scaling function $f(y) = \frac{1}{4} g''(y)$, which is proportional to the slope-slope correlation function, Pr{\"a}hofer and Spohn obtained exact numerical solutions for $f(y)$, which they found to behave as $f(y) \sim e^{-0.295 |y|^3}$ for large $y$, falling off faster than a Gaussian.

Evidence for anomalous transport in the Heisenberg spin chain at nonzero temperature was first found in the late 1990s. Sachdev and Damle \cite{SachdevDamle97, DamleSachdev98} explained diffusive ($z=2$) nonzero-temperature transport in the easy-axis $(\Delta > 1)$ XXZ model even though quasiparticles propagate ballistically, whereas other works \cite{Zotos99,Prosen2011} found ballistic ($z=1$) behavior at finite temperature in the easy-plane $(\Delta < 1)$ regime, suggesting anomalous transport at $\Delta = 1$.

The first numerical evidence for anomalous transport in the infinite-temperature $\Delta = 1$ Heisenberg spin chain was provided in 2011 by Ref.~\cite{Znidaric2011}. The $z=3/2$ exponent was demonstrated numerically in 2017 \cite{ljubotina2017spin}, in partially polarized domain wall initial states similar to those studied in our work, which approach the infinite-temperature state as $\mu\to 0$. The 3/2 exponent alone was not enough for the authors to propose that the Heisenberg spin chain is in the KPZ universality class, as it could have other explanations. In 2019 \cite{Ljubotina_PRL_2019}, however, they found numerically, for both the continuous-time Heisenberg model and the Floquet version studied here, that the two-point spin-spin correlation function at infinite temperature precisely matched the KPZ prediction for the slope-slope correlation function, Eq.~\eqref{eq:KPZ_correlation_function}, including the deviations from Gaussian at the tails. They therefore proposed that the infinite-temperature spin-1/2 Heisenberg model is in the KPZ universality class, with $\sigma_i^z \leftrightarrow \partial_x h(x_i)$, and that the infinite-temperature initial condition on the spin chain side corresponds to the stationary initial state (see Table \ref{tab:tableS1}) on the KPZ side. In their work, they used the finite-$\mu$ domain wall states studied here as a computational tool for obtaining the two-point correlation function at infinite temperature.

A number of works have proposed theoretical explanations for the observed $z=3/2$ dynamical exponent in the Heisenberg spin chain (e.g.,~\cite{Ilievski_PRL_2018, Sarang_PRL_2019, DeNardisPRL2019, gopalakrishnan2019anomalous, Vir2020_gauge, de2020superdiffusion}, see \cite{Bulchandani_StatMech_2021} for a review). The picture that emerges from these theoretical explanations is that the $z=3/2$ dynamical exponent and the two-point correlation function are universal for 1D integrable quantum systems with a non-Abelian global symmetry \cite{Ilievski_PRL_2018,  de2020superdiffusion, NormanYao22}, which, in the spin-1/2 Heisenberg model, is SU$(2)$.

The $z=3/2$ dynamical exponent has also been observed in several experiments \cite{wei2022quantum, scheie2021detection, keenan2022evidence}. Ref.~\cite{wei2022quantum}, for example, studied the transferred magnetization, which, assuming $\sigma_z \leftrightarrow \partial_x h$, corresponds to $2h(0,t) - h(\infty, t) - h(-\infty, t)$ for domain-wall initial states with several initial imbalances $\mu$. They confirmed the KPZ predictions that the mean and variance of the transferred magnetization both grow as $t^{2/3}$. They also measured the skewness which, in a nonzero-$\mu$ domain wall initial state, is expected to asymptote to 0.2241 (see Table \ref{tab:moments}). They measured $0.33 \pm 0.08$, where the uncertainty is one standard deviation, a result consistent with the KPZ prediction. They also confirmed that breaking either integrability or SU$(2)$ symmetry causes the dynamics to become either ballistic ($z=1$) or diffusive $(z=2)$. However, they measured the skewness for a domain wall with a very large initial imbalance ($\mu=1.5$), whereas the KPZ dynamics are expected to emerge at small $\mu$.

However, there is a problem with the conjecture that the infinite-temperature ($\mu=0$) Heisenberg model is in the KPZ universality class. As Refs.~\cite{Krajnik_PRL_2022, denardis2023nonlinear} point out, the probability distribution of the transferred magentization in this state must be symmetric; excitations are just as likely to move from the right side of the chain to the left as from the left to the right. Therefore, all of the odd moments of this distribution must be zero. This differs from the Baik-Rains distribution, which has a nonzero skewness of 0.359 \cite{PrahoferSpohn2000}; see Table \ref{tab:moments}.

If one reversed the order of limits, first taking $t\to\infty$ and then $\mu\to0$, the resulting transferred magnetization distribution may be skewed because even an infinitesimal domain wall breaks the mirror symmetry of the $\mu=0$ state. However, this does not resolve the issue because the late-time behavior at nonzero $\mu$ has been shown to be diffusive rather than KPZ \cite{Krajnik_PRL_2022, Gopalakrishnan_2023}. Ref.~\cite{Vir2020_gauge} would suggest otherwise, namely, that KPZ dynamics emerge even for $\mu>0$. In that work, the Heisenberg spin chain is coarse-grained and the global SU$(2)$ symmetry is promoted to a gauge symmetry, with a dynamical gauge field specifying the direction of the local Bethe vacuum in each lattice cell. A long-wavelength torsional mode of the gauge field is shown to obey a stochastic Burgers equation when the quasiparticle occupancy is uniform across cells, a condition that also holds  for the $\mu>0$ domain-wall states. However, this work does not connect the dynamics of the torsional mode with the dynamics of $\sigma_i^z$, the variable that is has been observed to play the role of $\partial_x h$. A summary of the different regimes discussed here and ways in which they differ from the KPZ universality class is shown in Table \ref{tab:comparison}.

\begin{figure}[h!]
    \centering
    \includegraphics{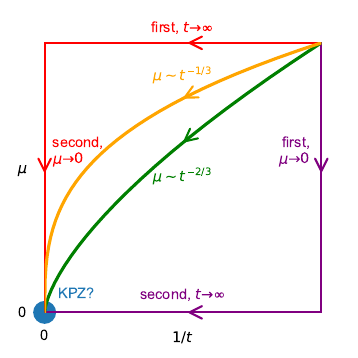}
    \caption{Different ways of taking the limits $t\to\infty$ and $\mu \to 0$, i.e.~of approaching the point where KPZ is conjectured. Taking the limit along any curve above the orange curve results in diffusive dynamics \cite{Gopalakrishnan_2023}. This includes the red curve, which corresponds to taking time to infinity first. The purple curve results in a symmetrically-distributed transferred magnetization, differing from the KPZ prediction, which is skewed \cite{denardis2023nonlinear, krajnik2023universal, Krajnik_PRL_2022}. The green curve shows an example of a way to take the limits simultaneously. Although this way could exhibit a skewness and a dynamical exponent consistent with KPZ, its predicted kurtosis disagrees with our observations for the times considered in our experiment.}
    \label{fig:order_of_limits}
\end{figure}

There are other ways of taking the limit, illustrated in Fig.~\ref{fig:order_of_limits}. So far, we have considered taking $\mu\to0$ first (purple line) or $t\to\infty$ first (red line), neither of which can result in KPZ dynamics. Taking a simultaneous limit can avoid these theoretical arguments. In particular, diffusive dynamics are expected to emerge at a time that scales as $1/\mu^3$ \cite{Gopalakrishnan_2023}. Therefore, if $t$ is scaled with $\mu$ in such a way that it remains less than $1/\mu^3$, for infinitesimal $\mu$, but still approaches infinity, the theoretical arguments against KPZ are avoided and the experiment has a chance of providing new information. In Fig.~\ref{fig:order_of_limits}, the orange curve, $\mu \sim t^{-1/3}$ indicates the threshold for diffusive dynamics; any curve approaching the origin that remains above the orange curve will result in diffusion. The purple curve results in symmetrically-distributed transferred magnetization, unlike the KPZ prediction, which is skewed. The green curve is an example of a way of taking the limit that is not ruled out theoretically. Although there is no prior theoretical motivation for taking the limit in any order other than $\mu\to0$ first, the collapsing behavior observed in Fig.~\ref{fig:Fig3_numerics} suggests that the skewness may be a function of $\mu t^{2/3}$, which is constant along the $\mu \sim t^{-2/3}$ curve. Further, from Fig.~\ref{fig:Fig3_numerics} and Fig.~3 of the main text, if $\mu t^{2/3}$ is fixed to a large number (at least about 10), the skewness appears to be consistent with that of the TW GUE probability distribution. However, we experimentally find a kurtosis of about $-0.05\pm 0.02$ (Fig.~3 of the main text and Fig.~\ref{fig:Fig3_numerics}), inconsistent with the TW GUE kurtosis of 0.09. This rules out KPZ dynamics on the timescale of the experiment. It does not rule out KPZ dynamics at much later times, but we do not see evidence for these dynamics either.

Therefore, the challenge is to explain why the dynamical exponent and two-point correlation function, at infinite temperature, look like KPZ, universally across integrable 1D quantum systems with a global non-Abelian symmetry, and yet other observables, such as the skewness of transferred magnetization when $\mu=0$, differ. In response to this challenge, Refs.~\cite{Krajnik_PRL_2022, krajnik2023universal} proposed and studied a classical Landau-Lifshitz (CLL) model and Refs.~\cite{Vir2020_gauge,denardis2023nonlinear} a non-linear fluctuating hydrodynamics (NLFH) model. Both systems predict a symmetric distribution for the transferred magnetization. CLL predicts an excess kurtosis close to 0, whereas NLFH model predicts 0.14. Although CLL agrees nicely with the findings reported in this work, it is an example of another system with similar behavior rather than an explanation of why these features should appear in the Heisenberg spin chain.

{
\setlength{\tabcolsep}{0mm}
\begin{table}[h!]\centering
\scriptsize
\begin{tabular}{c<{\hspace{2mm}}c<{\hspace{2mm}}cc<{\hspace{2mm}}c<{\hspace{2mm}}c<{\hspace{2mm}}c<{\hspace{2mm}}ccc}\toprule
&SU$(2)$? &Dynamical exponent &Skewness in Heisenberg &\multicolumn{3}{c}{Skewness in KPZ} & Universality of Heisenberg \\
& & & &$h(x,t=0)$ &$p\left(\mathcal{M}\right)$ &$\mathcal{S}$ \\\midrule
\makecell{$\mu = \infty$ (pure DW),\\ $t$ finite} &No &\begin{tabular}{l<{\hspace{2mm}}c<{\hspace{2mm}}}
Source &$z$ \\\midrule
Numerics \cite{ljubotina2017spin, Ljubotina_2017} & 5/3\\
Fig.~\ref{fig:muinf}C & 5/3\\
\end{tabular}
& $\sim 0.3$ (Fig.~\ref{fig:muinf_zoomed}A)
&
Wedge &TW GUE &0.224 & \makecell{KPZ not conjectured:\\$t$ finite, SU$(2)$ broken} \vspace{2mm} \\
\rowcolor{Gray}
$\mu = \infty$, $t\to\infty$ &No & \begin{tabular}{c} log-enhanced diffusion \\ ($z=2$) \cite{Misguich2017, stephan2017return} \end{tabular}
& 
&Wedge &TW GUE &0.224 & \begin{tabular}{c}KPZ not conjectured:\\SU$(2)$ broken \end{tabular}\vspace{2mm}\\
$\mu$ finite, $t$ finite &No &  \begin{tabular}{l<{\hspace{2mm}}c<{\hspace{2mm}}c<{\hspace{2mm}}c}
Source &$\mu$ &$z$ \\\midrule
Ref.~\cite{wei2022quantum} &0.22 &1.54 \\
Ref.~\cite{wei2022quantum} &1.5 &1.49 \\
Ref.~\cite{wei2022quantum} &1.8 &1.45 \\
Fig.~\ref{fig:S7}, sim &0.5 &1.49 \\
Fig.~\ref{fig:S7}, exp &0.5 &1.59 \\
Fig.~\ref{fig:S7}, sim &1.0 &1.59 \\
Fig.~\ref{fig:S7}, exp &1.0 &1.75 \\
\end{tabular} &\makecell{Ref.~\cite{wei2022quantum} ($\mu = 1.5$): 0.3\\This work: function of $\mu t^{2/3}$ \\Consistent with Ref.~\cite{wei2022quantum}.} &Wedge &TW GUE &0.224 & \makecell{ KPZ not conjectured: \\ $t$ finite, SU$(2)$ broken } \vspace{2mm} \\
\rowcolor{Gray}
$t \to \infty$, then $\mu \to 0$ &Subtle$^*$ &\begin{tabular}{l<{\hspace{2mm}}c<{\hspace{2mm}}c}
Source &$z$ \\\midrule
Numerics \cite{Krajnik_PRL_2022} &2 \\
Theory \cite{Gopalakrishnan_2023} & 2  \\

\end{tabular}& \begin{tabular}{c} Consistent with 0.224\\(Figs 3, \ref{fig:Fig3_numerics} with $\mu t^{2/3} \to \infty$)\end{tabular} &Wedge &TW GUE &0.224& Not KPZ: Wrong $z$ \vspace{2mm}\\
\makecell{$\mu = 0$ ($\infty$-temp state),\\
$t\to\infty$\\ ($\mu \to 0$, then $t\to\infty$)} &Yes &\begin{tabular}{l<{\hspace{2mm}}c<{\hspace{2mm}}c}
Source &$1/(2\beta)$ \\\midrule
Numerics \cite{ljubotina2017spin, Ljubotina_PRL_2019} &3/2 \\
Theory \cite{Ilievski_PRL_2018, Sarang_PRL_2019, DeNardisPRL2019, gopalakrishnan2019anomalous, Vir2020_gauge, de2020superdiffusion} & 3/2  \\
Fig.~\ref{fig:S7}B, sim &1.4\\
Fig.~\ref{fig:S7}B, exp &1.6\\
\end{tabular} &\makecell{Zero by symmetry }&Stationary &Baik-Rains &0.359 & Not KPZ: Wrong $\mathcal{S}$
\end{tabular}
\caption{Regimes of the 1D spin-1/2 Heisenberg model, starting in $\mu$-parametrized domain wall states (Eq.~3 of the main text). From the left, the first column indicates the regime, i.e.,~the initial imbalance $\mu$ and the time $t$. The second column indicates whether the initial state is SU$(2)$-symmetric. The third column lists the dynamical exponent in this regime. The fourth column indicates the skewness of the transferred magnetization. Column 5 shows the corresponding initial condition of the KPZ height function, assuming that $\sigma_z \sim \partial_x h$. Column 6 states the asymptotic probability distribution of the regularized KPZ height function for the given initial condition (see Table \ref{tab:moments}); the regularized height function corresponds to the transferred magnetization if $\sigma_z \sim \partial_x h$. The seventh column states the skewness of the probability distribution in the sixth column (see Table \ref{tab:moments}). Finally, the eighth column lists reasons why the particular case is not in the KPZ universality class. In the second-to-last row, ``subtle" refers to the fact that a state with infinitesimal $\mu$ breaks the SU$(2)$ symmetry by an infinitesimal amount. Such states would be appropriate for studying spontaneous symmetry breaking, where the symmetry might be broken by an infinitesimal external field. }
\label{tab:comparison}
\end{table}
}

\clearpage
\section{Unifying fSim conventions}
\subsection{Placement of the phase angle}
In our work, we use the following definition of the fSim gate:
\begin{equation}
U_{\mathrm{fSim}}=\left(\begin{matrix}
1 & 0 & 0 & 0\\
0 & \cos(\theta) & i\sin(\theta) & 0\\
0 & i\sin(\theta) & \cos(\theta) & 0\\
0 & 0 & 0 & e^{-i\phi}
\end{matrix}\right),
\end{equation}
which is a fully general number-conserving two-qubit gate up to single-qubit $Z$ rotations. Another natural choice would be to split the phase between the $|00\rangle$ and $|11\rangle$ states:
\begin{equation}
\label{eq:fsim_alternate}
U_{\mathrm{fSim}}'=\left(\begin{matrix}
e^{-i\phi/2} & 0 & 0 & 0\\
0 & \cos(\theta) & i\sin(\theta) & 0\\
0 & i\sin(\theta) & \cos(\theta) & 0\\
0 & 0 & 0 & e^{-i\phi/2}
\end{matrix}\right).
\end{equation}
This latter definition, which is more directly related to the trotterized Heisenberg Hamiltonian, is related to ours by
\begin{equation}
    U'_{\rm fSim} = e^{-i \sigma_1^z\phi/4} e^{-i \sigma_2^z\phi/4} U_{\rm fSim} =  U_{\rm fSim} e^{-i \sigma_1^z\phi/4} e^{-i \sigma_2^z\phi/4}.
\end{equation}
It is not immediately obvious that our experiment is insensitive to whether we use $U_{\rm fSim}$ or $U'_{\rm fSim}$, but this turns out to be the case. The transferred magnetization is independent of whether one uses $U'_{\rm fSim}$ or $U_{\rm fSim}$ as long as the number of cycles $t$ is at most $N_Q/2$. We verified this similarly to how we verified that the transferred magnetization is independent of $N_Q$ as long as $t \leq N_Q/2$ in Table \ref{tab:tableS1}. For example, with $\mu=0.5$, and $(\theta, \phi) = (0.4\pi, 0.8\pi)$, a density matrix simulation on four qubits gives a kurtosis after two cycles of $-0.30867052$, to this many digits, regardless of whether one uses $U_{\rm fSim}$ or $U'_{\rm fSim}$, and an 8-qubit simulation gives a kurtosis of $-0.12588028$ at cycle 4 regardless of which fSim gate one uses. The choice matters starting at cycle $N_Q/2+1$, but we are not interested in times past $N_Q/2$ because we have already seen (Table \ref{tab:tableS1}) that finite size effects appear there. 

Similarly, the transferred magnetization is unchanged, even past time $N_Q/2$, under either $\theta\to-\theta$ or $\phi\to-\phi$.

\subsection{Comparison to the $\eta$, $\lambda$ parameterization}
Following Ref. \cite{Ljubotina_Ballistic_2019}, it can be shown that Floquet application of the gate unitary,
\begin{equation}
\left(\begin{matrix}
1 & 0 & 0 & 0\\
0 & \frac{\sin\eta}{\sin(\lambda+\eta)} & \frac{\sin\lambda}{\sin(\lambda+\eta)} & 0\\
0 & \frac{\sin\lambda}{\sin(\lambda+\eta)} & \frac{\sin\eta}{\sin(\lambda+\eta)} & 0\\
0 & 0 & 0 & 1
\end{matrix}\right),
\end{equation}
gives the desired Heisenberg Hamiltonian evolution with $\Delta=\cos(\eta)$ in the limit $\lambda\rightarrow 0$. Here, $\lambda$ is imaginary and $\eta$ is real in the gapless ($\Delta<1$) regime, while $\lambda$ is real and $\eta$ is imaginary in the gapped ($\Delta>1$) regime. We will here consider the latter case; however, a similar derivation applies for $\Delta<1$ as well.

Setting this equal to $e^{i\phi/2} U_{\rm fSim}'$ (Eq.~\ref{eq:fsim_alternate}), we require that:
\begin{equation}
    ie^{i\phi/2}\sin(\theta)=\frac{\sin\lambda}{\sin(\lambda+\eta)}
\end{equation}
Comparing the magnitudes and phases of the two sides of this equation, one finds, respectively:
\begin{equation}
    \begin{split}
    \tan^2(\theta)=-\frac{\sin^2(\lambda)}{\sin^2(\eta)},\\
    \tan\left(\frac{\phi}{2}\right)=\frac{i\tan (\lambda)}{\tan(\eta)}.
    \end{split}
\end{equation}
Eliminating $\lambda$ and using $\Delta=\cos(\eta)$, we have:
\begin{align}
    \Delta^2\tan^2(\theta)&=\tan^2(\phi/2)\left(1+(1-\Delta^2)\tan^2(\theta)\right),\\
    \Delta&=\frac{\sin(\phi/2)}{\sin(\theta)}
\end{align}


\begin{thebibliography}{10}

\bibitem{KADANOFF1990}
L.~P. Kadanoff, {\it Physica A Stat. Mech. Appl.\/} {\bf 163}, 1 (1990).

\bibitem{SachdevBook}
S.~Sachdev, {\it Quantum Phase Transitions\/} (Cambridge University Press, 2011).

\bibitem{Geza}
G.~Odor, {\it Universality In Nonequilibrium Lattice Systems: Theoretical Foundations\/} (World Scientific publishing Co., 2008).

\bibitem{PolkovnikovRMP2011}
A.~Polkovnikov, {\it et~al.\/}, {\it Rev. Mod. Phys.\/} {\bf 83}, 863 (2011).

\bibitem{gogolin2016equilibration}
C.~Gogolin, J.~Eisert, {\it Rep. Prog. Phys.\/} {\bf 79}, 056001 (2016).

\bibitem{Zwierlein2011}
A.~Sommer, {\it et~al.\/}, {\it Nature\/} {\bf 472}, 201 (2011).

\bibitem{erne2018universal}
S.~Erne, R.~B{\"u}cker, T.~Gasenzer, J.~Berges, J.~Schmiedmayer, {\it Nature\/} {\bf 563}, 225 (2018).

\bibitem{joshi2022observing}
M.~K. Joshi, {\it et~al.\/}, {\it Science\/} {\bf 376}, 720 (2022).

\bibitem{zu2021emergent}
C.~Zu, {\it et~al.\/}, {\it Nature\/} {\bf 597}, 45 (2021).

\bibitem{wei2022quantum}
D.~Wei, {\it et~al.\/}, {\it Science\/} {\bf 376}, 716 (2022).

\bibitem{fontaine2021observation}
Q.~Fontaine, {\it et~al.\/}, {\it Nature\/} {\bf 608}, 687 (2022).

\bibitem{MitraPRA2014}
J.~Lux, {\it et~al.\/}, {\it Phys. Rev. A\/} {\bf 89}, 053608 (2014).

\bibitem{Vir_PRB2018}
V.~B. Bulchandani, R.~Vasseur, C.~Karrasch, J.~E. Moore, {\it Phys. Rev. B\/} {\bf 97}, 045407 (2018).

\bibitem{OlallaPRX2016}
O.~A. Castro-Alvaredo, B.~Doyon, T.~Yoshimura, {\it Phys. Rev. X\/} {\bf 6}, 041065 (2016).

\bibitem{Bertini_PRL_2016}
B.~Bertini, M.~Collura, J.~De~Nardis, M.~Fagotti, {\it Phys. Rev. Lett.\/} {\bf 117}, 207201 (2016).

\bibitem{DoyonSciPostPhys}
B.~Doyon, T.~Yoshimura, {\it SciPost Phys.\/} {\bf 2}, 014 (2017).

\bibitem{BulchandaniPRL2017}
V.~B. Bulchandani, R.~Vasseur, C.~Karrasch, J.~Moore, {\it Phys. Rev. Lett.\/} {\bf 119}, 220604 (2017).

\bibitem{de2018hydrodynamic}
J.~De~Nardis, D.~Bernard, B.~Doyon, {\it Phys. Rev. Lett\/} {\bf 121}, 160603 (2018).

\bibitem{Popper}
K.~Popper, {\it Conjectures and Refutations: The Growth of Scientific Knowledge\/} (Routledge Classics, 2002).

\bibitem{supremacy2019}
F.~Arute, {\it et~al.\/}, {\it Nature\/} {\bf 574}, 505 (2019).

\bibitem{Pan_PRL_2021}
Y.~Wu, {\it et~al.\/}, {\it Phys. Rev. Lett.\/} {\bf 127}, 180501 (2021).

\bibitem{Znidaric2011}
M.~\v{Z}nidari\v{c}, {\it Phys. Rev. Lett.\/} {\bf 106}, 220601 (2011).

\bibitem{ljubotina2017spin}
M.~Ljubotina, M.~{\v{Z}}nidari{\v{c}}, T.~Prosen, {\it Nat. Commun.\/} {\bf 8}, 1 (2017).

\bibitem{stephan2017return}
J.-M. St{\'e}phan, {\it J. Stat. Mech.: Theory Exp.\/} {\bf 2017}, 103108 (2017).

\bibitem{sanchez2018anomalous}
R.~J. S{\'a}nchez, V.~K. Varma, V.~Oganesyan, {\it Phys. Rev. B\/} {\bf 98}, 054415 (2018).

\bibitem{Ilievski_PRL_2018}
E.~Ilievski, J.~De~Nardis, M.~Medenjak, T.~Prosen, {\it Phys. Rev. Lett.\/} {\bf 121}, 230602 (2018).

\bibitem{Sarang_PRL_2019}
S.~Gopalakrishnan, R.~Vasseur, {\it Phys. Rev. Lett.\/} {\bf 122}, 127202 (2019).

\bibitem{DeNardisPRL2019}
J.~De~Nardis, M.~Medenjak, C.~Karrasch, E.~Ilievski, {\it Phys. Rev. Lett.\/} {\bf 123}, 186601 (2019).

\bibitem{gopalakrishnan2019anomalous}
S.~Gopalakrishnan, R.~Vasseur, B.~Ware, {\it Proc. Natl. Acad. Sci. U.S.A.\/} {\bf 116}, 16250 (2019).

\bibitem{de2020superdiffusion}
J.~De~Nardis, S.~Gopalakrishnan, E.~Ilievski, R.~Vasseur, {\it Phys. Rev. Lett\/} {\bf 125}, 070601 (2020).

\bibitem{MoorePRB2020}
M.~Dupont, J.~E. Moore, {\it Phys. Rev. B\/} {\bf 101}, 121106 (2020).

\bibitem{Dupont_PRL_2021}
M.~Dupont, N.~E. Sherman, J.~E. Moore, {\it Phys. Rev. Lett.\/} {\bf 127}, 107201 (2021).

\bibitem{IlievskiPhysRevX2021}
E.~Ilievski, J.~De~Nardis, S.~Gopalakrishnan, R.~Vasseur, B.~Ware, {\it Phys. Rev. X\/} {\bf 11}, 031023 (2021).

\bibitem{scheie2021detection}
A.~Scheie, {\it et~al.\/}, {\it Nat. Phys.\/} {\bf 17}, 726 (2021).

\bibitem{jepsen2020spin}
P.~N. Jepsen, {\it et~al.\/}, {\it Nature\/} {\bf 588}, 403 (2020).

\bibitem{Heisenberg}
W.~Heisenberg, {\it Z. Phys.\/} {\bf 49}, 619 (1928).

\bibitem{Neill2021accurately}
C.~Neill, {\it et~al.\/}, {\it Nature\/} {\bf 594}, 508 (2021).

\bibitem{supplement}
Materials and methods are available as supplementary materials.

\bibitem{Bethe1931}
H.~Bethe, {\it Z. Phys.\/} {\bf 71} (1931).

\bibitem{Essler_PRL_2013}
J.-S. Caux, F.~H.~L. Essler, {\it Phys. Rev. Lett.\/} {\bf 110}, 257203 (2013).

\bibitem{ProsenTrotter}
M.~Vanicat, L.~Zadnik, T.~Prosen, {\it Phys. Rev. Lett.\/} {\bf 121}, 030606 (2018).

\bibitem{Integrability2}
P.~W. Claeys, J.~Herzog-Arbeitman, A.~Lamacraft, {\it SciPost Phys.\/} {\bf 12}, 007 (2022).

\bibitem{Ljubotina_Ballistic_2019}
M.~Ljubotina, L.~Zadnik, T.~Prosen, {\it Phys. Rev. Lett.\/} {\bf 122}, 150605 (2019).

\bibitem{Ljubotina_PRL_2019}
M.~Ljubotina, M.~Znidaric, T.~Prosen, {\it Phys. Rev. Lett.\/} {\bf 122}, 210602 (2019).

\bibitem{keenan2022evidence}
N.~Keenan, N.~F. Robertson, T.~Murphy, S.~Zhuk, J.~Goold, {\it npj Quantum Information\/} {\bf 9}, 72 (2023).

\bibitem{Zotos99}
X.~Zotos, {\it Phys. Rev. Lett.\/} {\bf 82}, 1764 (1999).

\bibitem{SachdevDamle97}
S.~Sachdev, K.~Damle, {\it Phys. Rev. Lett.\/} {\bf 78}, 943 (1997).

\bibitem{DamleSachdev98}
K.~Damle, S.~Sachdev, {\it Phys. Rev. B\/} {\bf 57}, 8307 (1998).

\bibitem{denardis2023nonlinear}
J.~De~Nardis, S.~Gopalakrishnan, R.~Vasseur, {\it Phys. Rev. Lett.\/} {\bf 131}, 197102 (2023).

\bibitem{krajnik2023universal}
{\v{Z}}.~Krajnik, J.~Schmidt, E.~Ilievski, T.~Prosen, {\it Phys. Rev. Lett.\/} {\bf 132}, 017101 (2024).

\bibitem{Prahofer2004}
M.~Pr{\"a}hofer, H.~Spohn, {\it J. Stat. Phys.\/} {\bf 115}, 255 (2004).

\bibitem{KIP2020}
{\v{Z}}.~Krajnik, E.~Ilievski, T.~Prosen, {\it SciPost Phys.\/} {\bf 9}, 038 (2020).

\bibitem{Norm_PRL_2022}
B.~Ye, F.~Machado, J.~Kemp, R.~B. Hutson, N.~Y. Yao, {\it Phys. Rev. Lett.\/} {\bf 129}, 230602 (2022).

\bibitem{KPZ}
M.~Kardar, G.~Parisi, Y.-C. Zhang, {\it Phys. Rev. Lett.\/} {\bf 56}, 889 (1986).

\bibitem{Krajnik_PRL_2022}
Z.~Krajnik, E.~Ilievski, T.~Prosen, {\it Phys. Rev. Lett.\/} {\bf 128}, 090604 (2022).

\bibitem{gopalakrishnan2022theory}
S.~Gopalakrishnan, A.~Morningstar, R.~Vasseur, V.~Khemani, {\it Phys. Rev. B\/} {\bf 109}, 024417 (2024).

\bibitem{PrahoferSpohn2000}
M.~Pr\"ahofer, H.~Spohn, {\it Phys. Rev. Lett.\/} {\bf 84}, 4882 (2000).

\bibitem{Gopalakrishnan_2023}
S.~Gopalakrishnan, R.~Vasseur, {\it Rep. Prog. Phys.\/} {\bf 86}, 036502 (2023).

\bibitem{Vir2020_gauge}
V.~B. Bulchandani, {\it Phys. Rev. B\/} {\bf 101}, 041411 (2020).

\bibitem{Sarang_PRB_2022}
V.~B. Bulchandani, D.~A. Huse, S.~Gopalakrishnan, {\it Phys. Rev. B\/} {\bf 105}, 214308 (2022).

\bibitem{Manas1}
A.~Das, M.~Kulkarni, H.~Spohn, A.~Dhar, {\it Phys. Rev. E\/} {\bf 100}, 042116 (2019).

\bibitem{Manas2}
D.~Roy, A.~Dhar, H.~Spohn, M.~Kulkarni, {\it Phys. Rev. B\/} {\bf 107}, L100413 (2023).

\bibitem{Giamarchi}
T.~Giamarchi, {\it Quantum Physics in One Dimension\/} (Oxford University Press, 2003).

\bibitem{Vidal2008}
G.~Vidal, {\it Phys. Rev. Lett.\/} {\bf 101}, 110501 (2008).

\bibitem{evenbly2009algorithms}
G.~Evenbly, G.~Vidal, {\it Phys. Rev. B\/} {\bf 79}, 144108 (2009).

\bibitem{rosenberg_2024_10045874}
E.~Rosenberg, T.~I. Andersen, Dynamics of magnetization at infinite temperature in a Heisenberg spin chain  (2024). Zenodo. \href{https://doi.org/10.5281/zenodo.10045873}{doi:10.5281/zenodo.10045873}

\bibitem{recirq}
E.~Rosenberg, Kardar-Parisi-Zhang Dynamics in the Heisenberg Spin Chain  (2023). ReCirq. \href{https://quantumai.google/cirq/experiments/kpz/kpz}{https://quantumai.google/cirq/experiments/kpz/kpz}

\bibitem{Brooks2020}
B.~Foxen, {\it et~al.\/}, {\it Phys. Rev. Lett.\/} {\bf 125}, 120504 (2020).

\bibitem{bound_states}
A.~Morvan, {\it et~al.\/}, {\it Nature\/} {\bf 612}, 240 (2022).

\bibitem{MIP}
J.~C. Hoke, {\it et~al.\/}, {\it Nature\/} {\bf 622}, 481 (2023).

\bibitem{Quenouille1949}
M.~H. Quenouille, {\it J. R. Stat. Soc. Ser. B Methodol.\/} {\bf 11}, 68 (1949).

\bibitem{Tukey1958}
J.~W. Tukey, {\it Ann. Math. Stat.\/} {\bf 29}, 614 (1958).

\bibitem{JackknifeReview}
B.~Efron, {\it The Jackknife, the Bootstrap and Other Resampling Plans\/} (Society for Industrial and Applied Mathematics, 1982).

\bibitem{Note1}
For a particular initial bitstring state, the transferred magnetization is only independent of system size when $N_Q \geq 4t-2$, as one would expect from the causality structure of the circuit, but after averaging over initial bitstrings (or equivalently in the mixed initial state), we find the transferred magnetization is exactly independent of system size as long as $N_Q \geq 2t$ sites.

\bibitem{TEBD}
G.~Vidal, {\it Phys. Rev. Lett\/} {\bf 91} (2003).

\bibitem{cuQuantum}
L.~Fang, {ahehn-nv}, hhbayraktar, {sam-stanwyck}, {NVIDIA}/cu{Q}uantum: cu{Q}uantum {P}ython v22.11.0.1 (2023). Zenodo. \href{https://doi.org/10.5281/zenodo.7523366}{doi:10.5281/zenodo.7523366}

\bibitem{qsim}
{Quantum AI team and collaborators}, qsim (2020).

\bibitem{Bulchandani_StatMech_2021}
V.~Bulchandani, {\it et~al.\/}, {\it J. Stat. Mech.: Theory Exp.\/} {\bf 2021}, 084001 (2021).

\bibitem{PlishkeRacz85}
M.~Plischke, Z.~R\'acz, {\it Phys. Rev. A\/} {\bf 32}, 3825 (1985).

\bibitem{Family_1985}
F.~Family, T.~Vicsek, {\it J. Phys. A: Math. Gen.\/} {\bf 18}, L75 (1985).

\bibitem{Johansson_2000}
K.~Johansson, {\it Commun. Math. Phys.\/} {\bf 209}, 437 (2000).

\bibitem{Hartmann_2018}
A.~K. Hartmann, P.~L. Doussal, S.~N. Majumdar, A.~Rosso, G.~Schehr, {\it EPL\/} {\bf 121}, 67004 (2018).

\bibitem{tracywidom}
Y.-Y. Mao, Tracy{W}idom (2021). \href{https://github.com/yymao/TracyWidom}{https://github.com/yymao/TracyWidom}

\bibitem{bejan2005largest}
A.~I. Bejan, Largest eigenvalues and sample covariance matrices. {T}racy-{W}idom and {P}ainleve {II}: Computational aspects and realization in {S}-{P}lus with applications, Master's thesis, The University of Warwick (2005).

\bibitem{TW_expansion}
G.~Borot, C.~Nadal, {\it Random Matrices: Theory and Applications\/} {\bf 01}, 1250006 (2012).

\bibitem{BaikRains}
J.~Baik, E.~M. Rains, {\it Journal of Statistical Physics\/} {\bf 100}, 523 (2000).

\bibitem{Prosen2011}
T.~Prosen, {\it Phys. Rev. Lett.\/} {\bf 106}, 217206 (2011).

\bibitem{NormanYao22}
B.~Ye, F.~Machado, J.~Kemp, R.~B. Hutson, N.~Y. Yao, {\it Phys. Rev. Lett.\/} {\bf 129}, 230602 (2022).

\bibitem{Ljubotina_2017}
M.~Ljubotina, M.~\v{Z}nidari\v{c}, T.~Prosen, {\it J. Phys. A: Math. Theor.\/} {\bf 50}, 475002 (2017).

\bibitem{Misguich2017}
G.~Misguich, K.~Mallick, P.~L. Krapivsky, {\it Phys. Rev. B\/} {\bf 96}, 195151 (2017).

\end{thebibliography}
\end{document}